\documentclass[lettersize,journal]{IEEEtran}
\usepackage{amsmath,amsfonts}
\usepackage{amsmath,amssymb}
\usepackage{algorithmic}
\usepackage{algorithm}
\usepackage{array}
\usepackage[caption=false,font=small,labelfont=sf,textfont=sf]{subfig}
\usepackage{mathtools}
\usepackage{textcomp}
\usepackage{stfloats}
\usepackage{url}
\usepackage{cite}
\usepackage{hyperref}
\usepackage{verbatim}
\usepackage{graphicx}
\usepackage[shortlabels]{enumitem}
\usepackage[center]{caption}
\usepackage{subfig}
\usepackage{dsfont}
\usepackage{xcolor}
\def\BibTeX{{\rm B\kern-.05em{\sc i\kern-.025em b}\kern-.08em
		T\kern-.1667em\lower.7ex\hbox{E}\kern-.125emX}}
\usepackage{balance}
\DeclareMathOperator*{\argmin}{arg\,min}

\begin{document}
	\setlength{\abovecaptionskip}{3pt}
	\setlength{\belowcaptionskip}{-7pt} 
	
	\title{Version Age-Optimal Cached Status Updates in a Gossiping Network with Energy Harvesting Sensor}
	
	\author{Erfan Delfani, and Nikolaos Pappas, 
		\IEEEmembership{Senior Member, IEEE}.
		\thanks{The authors are with the Department of Computer and Information Science at Linköping University, Sweden, email: \{\texttt{erfan.delfani, nikolaos.pappas\}@liu.se}. This work has been supported in part by the Swedish Research Council (VR), ELLIIT, Zenith, and the European Union (ETHER, 101096526). A shorter version has been published in \cite{WiOpt23}.}}
	
	\markboth{}%
	{Version Age-Optimal Cached Status Updates in a Gossiping Network with Energy Harvesting Sensor}
	
	\maketitle
	
	\begin{abstract}
		In this work, we consider a real-time IoT monitoring system in which an energy harvesting sensor with a finite-size battery measures a physical process and transmits the status updates to an aggregator. The aggregator, equipped with caching capabilities, can serve the external requests of a destination network with either a stored update or a fresh update from the sensor. We assume the destination network acts as a gossiping network in which the update packets are forwarded among the nodes in a randomized setting. We utilize the Markov Decision Process framework to model and optimize the network's average Version Age of Information (AoI) and obtain the optimal policy at the aggregator. The structure of the optimal policy is analytically demonstrated and numerically verified. Numerical results highlight the effect of the system parameters on the average Version AoI. The simulations reveal the superior performance of the optimal policy compared to a set of baseline policies.
	\end{abstract}
	
	
	\section{Introduction}
	The freshness of information has become critical in many IoT-enabled systems used for real-time monitoring applications such as health care, transportation, and natural environment monitoring \cite{abd2019role, shreedhar2019age}. In these IoT systems, the sensors measure the state of physical processes, and the aggregated data in the source nodes are transmitted through a communication network to the destination nodes (monitors) for further processing. The processed data can be used for different purposes, like remote control of actuators in industrial automation applications or steering autonomous vehicles in intelligent transportation systems. Since the accuracy and reliability of such time-sensitive controls are highly related to the freshness of the received data (update packets) at the destination nodes, delivering fresh updates can be of crucial importance \cite{zhang2015mission}. Age of information (AoI) is a performance metric that quantifies the freshness of information defined as the time elapsed since the generation time of the last successfully received update at the monitor 
	\cite{kaul2012real, pappas2023age}. However, AoI fails to capture the content of information; thus, other content-aware metrics such as the age of incorrect information \cite{maatouk2020age}, age of synchronization \cite{zhong2018two}, age of outdated information \cite{liu20222}, or extended AoI metric \cite{stamatakis2019control} have been proposed in the literature of data freshness. Recently, a new content-based metric called Version AoI has been introduced, which measures how many versions out-of-date the information at a destination node is, compared to the version at the source \cite{yates2021age}. Version AoI offers the benefit of requiring minimal knowledge about the information source, primarily focusing on the changes occurring in the source's content, making it an appropriate metric in various practical applications. Version AoI has been studied in gossiping networks where the update packets are randomly forwarded among the network nodes \cite{yates2021age,yates2021timely,buyukates2022version,kaswan2022timely,kaswan2022susceptibility,mitra2022asuman,kaswan10051981,mitra2023age,mitra2023learning,abd2023distribution}.
	
	In practical IoT applications with battery-powered sensors, there is a severe limitation against frequent updates because each update consumes energy, and the batteries have a finite capacity. In the current IoT systems, a class of energy harvesting (EH) sensors with rechargeable batteries are being deployed and promise cost-efficient and self-sustainable IoT networks \cite{gunduz2014designing}. In this case, the intermittent nature of harvested energy requires more intelligent energy budget management while ensuring timely updates. 
	
	Caching is another important aspect of practical IoT networks for managing network resources and enhancing network performance. Caching involves storing and delivering frequently accessed information or content at the edge of the network, closer to sensors or devices. Effectively, it can reduce the need to retrieve data from the source, which can be energy-consuming and less efficient.
	
	In this work, we consider a model for a real-time IoT network in order to analyze and optimize the entire information process, spanning from the generation of information at the source to its delivery to the destination nodes. This model incorporates the integration of \textit{energy harvesting}, \textit{caching}, and \textit{gossiping} aspects within the network. Moreover, we consider the \textit{Version AoI} as the network's key performance metric to be optimized. We consider a general IoT monitoring setup comprising an EH sensor and an aggregator located near the sensor. The objective is to minimize the average Version AoI within a destination gossiping network while adhering to an energy budget and responding to requests from destination nodes within the network. The aggregator can cache the most recent update received and must decide whether to transmit a fresh or stored update upon request. To model and optimize the problem, we utilize the Markov Decision Process (MDP) framework. The structure of the optimal solution and the impact of various system parameters are investigated through analytical and numerical approaches.
	
	\subsection{Related Works}
	Several works have investigated the optimization of data freshness in energy-constrained IoT networks \cite{bacinoglu2015age, yates2015lazy, bacinoglu2017scheduling, sun2017update, bacinoglu2018achieving, arafa2018age, farazi2018average, feng2018minimizing, chen2019age, feng2021age, arafa2019age, bacinoglu2019optimal, abd2019online, abd2020reinforcement, leng2019minimizing, ceran2019reinforcement, gindullina2021age, stamatakis2019control, hatami2022demand, hatami2021aoi, holm2021freshness}. Our study primarily focuses on discrete-time systems that adopt the Markov Decision Process (MDP) framework to minimize the Age of Information (AoI) in energy harvesting networks. These works can be classified into two groups: \textit{permanent query} systems and \textit{request-based} systems.
	
	In the first group, the system's action does not depend on external requests or queries from the server or destination nodes, making them \textit{permanent query} systems. One such work is presented in \cite{stamatakis2019control}, where the authors investigate a system employing an energy harvesting sensor to monitor a two-state stochastic process and transmit status updates to a destination node. The cost function of the MDP problem incorporates two AoI variables with different exponents, considering higher demand in the alarm state. The optimal status updating policy is determined using the Value Iteration algorithm.
	In \cite{abd2019online}, the authors consider an IoT-enabled monitoring system in which an RF energy harvesting device samples and sends update packets to a destination node. They formulate an MDP problem to find the optimal online status updating policy that minimizes the long-term average AoI. The optimal policy accounts for battery dynamics, AoI, and channel state information, determining the allocation of each time slot for energy harvesting or status updating. The study also presents analytical results on the value function and optimal policy structure and compares age-optimal and throughput-optimal policies.
	Building upon the previous work, \cite{abd2020reinforcement} extends the system to multiple source nodes. The optimal policy aims to minimize the long-term average weighted sum of AoI values associated with different physical processes monitored by the source nodes. Additionally, the authors propose a computationally-efficient deep reinforcement learning (DRL) algorithm to learn the age-optimal policy.
	In \cite{leng2019minimizing}, the focus shifts to a cognitive radio system where the secondary user is an energy harvesting sensor. The sensor decides between spectrum sensing and status updating operations at each time slot. The sequential decision-making problem is modeled as a Partially Observable Markov Decision Process (POMDP) and solved using dynamic programming. The study also explores the structural properties of the optimal policy.
	Considering a status update system over an error-prone channel, \cite{ceran2019reinforcement} investigates an energy harvesting transmitter (TX) that can sample and transmit a new update, remain idle, or retransmit the last update to a receiver (RX). An average-cost MDP problem is formulated, accounting for the energy consumption during sampling and transmission, as well as the limited battery capacity of the TX. The age-optimal scheduling policy is determined using the proposed Randomized Value Iteration (RVI) and Reinforcement Learning (RL) algorithms for known and unknown channel and energy harvesting statistics, respectively.
	In \cite{gindullina2021age}, the authors consider an energy harvesting monitoring node with a finite battery capacity. The node receives status updates from multiple heterogeneous information sources, each measuring an underlying process. These sources exhibit varying energy consumption and AoI values. The monitoring node aims to minimize the average AoI by selecting an optimal action at each time slot, including requesting an update from a source or remaining idle. The problem is formulated as an MDP, and the optimal solution structure is analyzed. The Value Iteration algorithm is employed for finding the optimal policy.
	
	Regarding the \textit{request-based} approach, \cite{hatami2022demand} focuses on minimizing the on-demand Age of Information subject to energy causality and transmission constraints in a multi-user and multi-sensor IoT energy harvesting sensing network. The problem is formulated as an MDP, and an iterative algorithm is proposed to obtain the optimal status updating policy at the cache-enabled edge node. Additionally, a sub-optimal low-complexity algorithm is developed to handle scenarios with a large number of sensors.
	In \cite{hatami2021aoi}, the authors address the problem without transmission constraints using an MDP framework. A model-free Q-learning method is proposed to search for an optimal policy in cases where transition probabilities are unknown.
	Lastly, in \cite{holm2021freshness}, a pull-based communication model is introduced, which assesses the freshness of status updates only when queries are made. The Age of Information at Query (QAoI) metric is employed as the cost function of an MDP problem. The optimal status updating policy is determined for a monitoring scenario where an edge node receives periodic queries from a server regarding the information of an energy harvesting sensor (with known periodicity).
	
	\subsection{Contributions}
	In the AoI literature, the MDP framework has been extensively employed. However, this framework has not been previously considered when studying Version AoI in a gossiping network scenario. Several works \cite{yates2021age,yates2021timely,buyukates2022version,kaswan2022timely,kaswan2022susceptibility,mitra2022asuman,kaswan10051981,mitra2023age} have focused on examining the growth or scaling of average Version AoI within permanent query gossiping networks of varying sizes, topologies, and with different adversarial factors such as timestomping and jamming. \cite{mitra2023learning} has proposed a learning-based approach to optimize update rates in sparse gossip networks to minimize the average Version age of the worst-performing node, using a continuum-armed bandit formulation of GP-UCB algorithm, while \cite{abd2023distribution} has introduced a method to characterize the higher-order statistics of Version AoI in gossiping networks with different topologies. In this work, we utilize the MDP framework to optimize the Version AoI in a request-based IoT monitoring setup consisting of an EH sensor that measures a physical process, a cache-enabled aggregator near the sensor, and a gossiping network for data transmission. The aggregator is responsible for deciding whether to send a fresh update or a previously stored one to nodes in the network when a request is received. The main contributions of this paper are summarized below:
	\begin{itemize}
		\item We utilize the MDP framework to formulate and optimize the average Version AoI of the gossiping network whose nodes are updated by the EH sensor through the actions adopted by the cache-enabled aggregator.
		\item We show that the MDP problem has an optimal solution. We further prove that this optimal policy is independent of the state of the network's nodes.
		\item We demonstrate that the optimal policy has a threshold structure based on the Version AoI at the aggregator.
		\item The analytical results are verified through simulations, and the impact of the system's parameters on the average Version AoI is highlighted.
	\end{itemize}
	
	\section{System Model and Problem Formulation}
	
	\label{SystemModel}
	\begin{figure}[!t]
		\centering
		\includegraphics[width=2.3in]{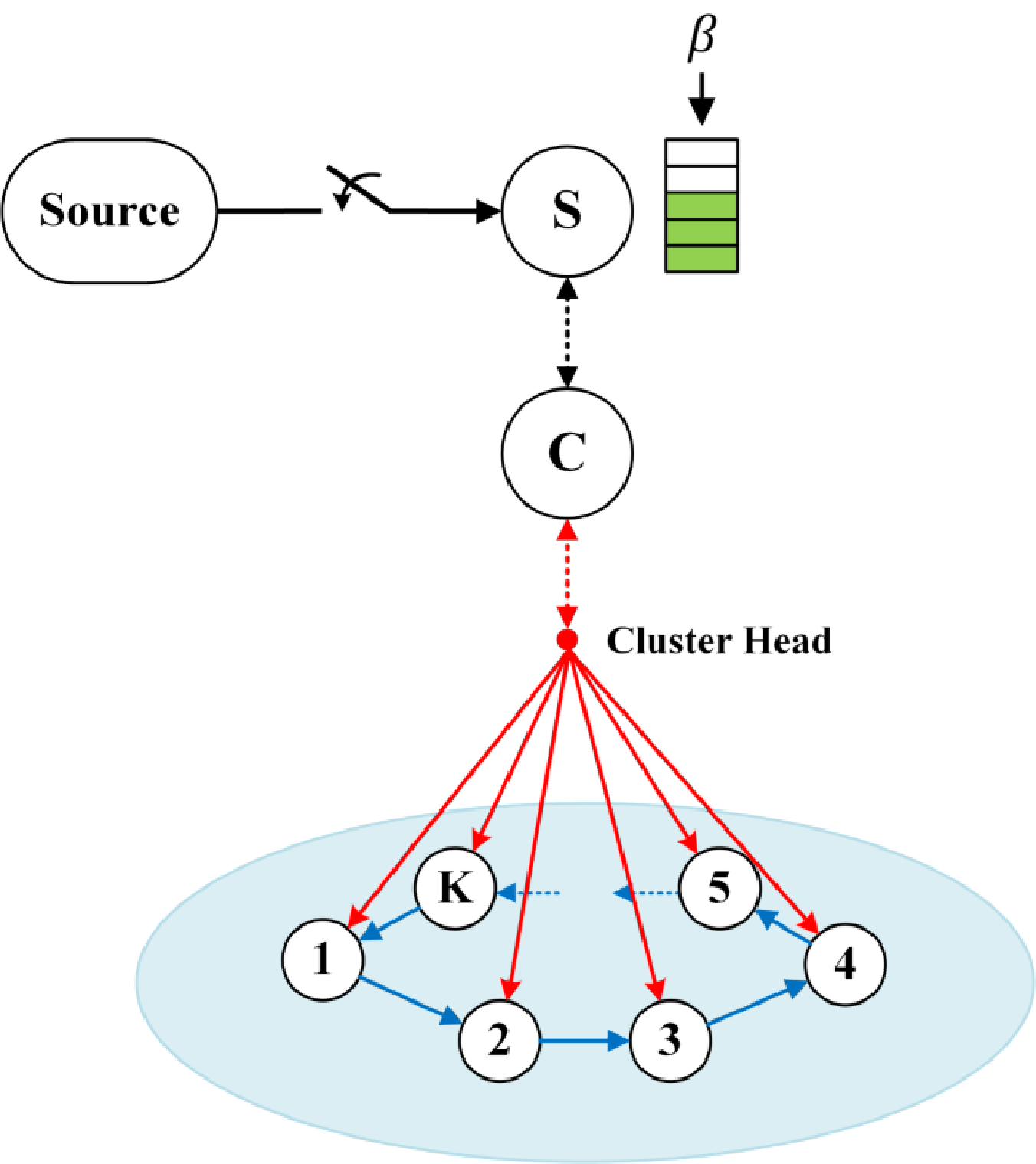}
		\caption{The considered system model.}
		\label{fig1}
	\end{figure}
	
	We consider a discrete-time monitoring system as in Fig. \ref{fig1}. This system comprises an energy harvesting sensor S, an aggregator C, and $K$ destination nodes. The sensor S can measure the source process and send the measurement in the form of a status update packet to the aggregator C. C can store (cache) the last received update, and it has to decide whether to request a fresh update from S or to serve the external request with a cached update. Destination nodes send external requests through a cluster head. Moreover, the destination nodes are connected, and each node can send its current stored status to its neighboring nodes by local gossiping. 
	
	The sensor S is equipped with a rechargeable battery of finite size $B$ and is harvesting energy from ambient sources. The energy arrival process is modeled as a Bernoulli process with average probability $0<\beta<1$. We assume that at each time slot, at most one request from one of the external nodes will be served by C. The service probability of external requests follows a categorical distribution (or generalized Bernoulli) where the external request from node $k\in\left\{1, 2,\ldots, K\right\}$ is served with probability $0<q_k<1$, and no external request is served with probability $q=1-\sum_{k=1}^{K}q_k$. When there is an external request from node $k$, if S has an empty battery, a cached update will be served by C. Otherwise, if the battery is not empty, C should decide whether to request a fresh update from S or to serve the node $k$ with a cached update. If there is no external request, then C will not ask for a fresh update from S. Every time C receives a fresh update, the new update is stored, and the previously cached one is dropped. In addition, when a fresh status update is requested from S, we assume that the overall duration of requesting, sampling, and transmission (through C) occupies one time slot and consumes one energy unit from the battery of S. The transmission of a cached packet from C to external nodes also takes one time slot and is assumed to be error-free. In this system, the state of the source changes with a probability $p_t>0$ at each time slot. In this general model, any gossiping topology can be considered for the destination network; however, in this work, we consider a uni-directional ring topology for the network where the node $k$ is updated by its neighboring node $(k-1)$ with the probability $\lambda_k$. Notice that node $1$ is updated by node $K$. We assume that gossiping is error-free and that each update transmission occupies one time slot.
	
	Version AoI at node $k$, denoted by $\Delta_k$, is defined as how many versions out-of-date the information at the node $k$ is, compared to the version at the source. In this regard, the source always maintains the current version of its status, and thus $\Delta_{Source}(t)\!=\!0, \forall t\!=\!0,1,\cdots$. When node $k$ is updated by the current state of the source at time slot $t$, and there is no state change at the source, then $\Delta_k(t\!+\!1)\!=\!0$. Otherwise, if the source state changes in time slot $t$, then $\Delta_k(t+1)\!=\!1$. The value of $\Delta_k(t+1)$ can be obtained as follows:
	\begin{align*}
		&\Delta_k(t+1)=\mathds{1}\!\left(\text{Source state changes in time slot $t$} \right) \\
		&+
		\begin{cases}
			0 & \parbox[t][][t]{4.6cm}{\small {Node $k$ receives fresh update via aggregator C (gossiping irrelevant),}} \\
			\Delta_C(t) & \parbox[t][][t]{4.6cm}{\small {Node $k$ receives cached update via C (gossiping irrelevant),}} \\
			\min\!\left\{\Delta_k(t),\Delta_{k\!-\!1}(t)\right\} & \parbox[t][][t]{4.6cm}{\small {Node $k$ is updated through gossiping (not via aggregator),}} \\
			\Delta_k(t) & \parbox[t][][t]{4.6cm}{\small {Node $k$ is not updated.}}
		\end{cases} \notag
	\end{align*}

	$\mathds{1}(\cdot)$ is the indicator function, it is equal to $1$ when its argument is true, and $0$ otherwise. We further assume that the Version age is upper bounded by $\Delta_{max}$ since the stored data with large Version age becomes too stale to be useful; thus, there is no point in having larger values of Version AoI.
	
	\subsection{Markov Decision Process (MDP) Formulation}
	
	Our objective is to model the average Version AoI for the updates stored at the destination nodes and find an optimal status update policy at the aggregator that minimizes the average Version AoI. The update policy $\pi$ is defined as a sequence of actions $\pi =\left(a^{\pi}(1),a^{\pi}(2),\dots \right)$ where $a^{\pi}\left(t\right)=1$ if the aggregator C requests for a fresh update from S and $a^{\pi }\left(t\right)=0$ if C serves the external request with a cached update. The action of the aggregator at each time slot depends on the state of the battery, the state change of the source, the arrival and service probability of external requests, and gossiping (are defined in \ref{TransProb}). By $\mathrm{\Pi}$ we denote the set of all causal policies; then, the problem can be formulated as an infinite horizon average cost MDP, consisting of a tuple $<\!S,A,P,C\!>$, where $S$ is the state space, $A$ is the set of actions, $P$ is the state transition probability function, and $C$ is the cost of MDP.
	
	\begin{itemize}
		
		\item \textbf{States}: The state of system at time slot $t$ is defined by $s(t)\!=\!\big(b(t),\Delta_1(t),{\Delta}_2(t),\dots,{\Delta}_K(t),{\Delta}_C(t)\big)$, where $b(t)\!\in\!\{0,1,\dots ,B\}$ is the state of the battery, ${\Delta}_k(t)\!\in\!\{0,1,\dots,\Delta_{max}\}$ is the Version AoI of node $k\!\in\!\{1,2,\dots,K\}$, and ${\Delta}_C(t)\!\in\!\{0,1,\dots,\Delta_{max}\}$ is the Version AoI at the aggregator. 
		
		It should be noted that the index of the requesting node can also be included in the state vector. However our analysis in Appendix \ref{AppenNewState} shows that the value function's structure remains consistent, yielding similar results presented in Section \ref{AnalyticalResults}.Thus, we choose to keep the state space smaller for the sake of simplicity and tractability.
		
		\item \textbf{Actions}: The action at time $t$, denoted by $a(t)$, indicates the decision of the aggregator C if requests for a new update from sensor S, $a(t)=1$, or serves the external request with a cached update, $a(t)=0$.
		\item \textbf{Transition probabilities}: The transition probabilities between the different states, presented in Section \ref{TransProb}.
		\item \textbf{Cost}: The instantaneous cost of MDP $C\big(s(t),a(t)\big)$ is equal to the average Version AoI ${\Delta}_{AVG}(t)=\frac{1}{K}\sum^K_{k=1}{{\Delta}_k(t)}$.
		
	\end{itemize}
	
	Our MDP Problem can be formulated as follows:
	\begin{align}
		\label{MDP_eqn}
		\mathop{\mathrm{min}}_{\pi \in \mathrm{\Pi }} {\mathop{{\mathrm{lim} \mathrm{sup}\ }}_{T\to \infty } \frac{1}{T}E\left[\sum^{T-1}_{t=0}{{\Delta}^{\pi }_{AVG}(t)}\Big|s(0)=s_0\right]},
	\end{align}
	where $\pi$ is a policy that defines action $a(t)$ at each time slot based on the system's current state. 
	
	\subsection{Transition Probabilities}
	\label{TransProb}
	
	In this section, we provide the transition probabilities among MDP problem states by introducing the following random variables.
	
	\textit{Energy arrival process} is denoted by ${\left\{{\mathcal{\textit{e}}}(t)\right\}}^{\infty}_{t=0}$, where ${\mathcal{\textit{e}}}(t)\in \left\{0,1\right\}$:
	\begin{align}
		P\left[\mathcal{\textit{e}}(t)=e\right]=
		\begin{cases}
			\beta & e=1, \\
			1-\beta & e=0, \\
			0 & \text{otherwise.}
		\end{cases}
	\end{align}
	
	\textit{Source state change process} is indicated by ${\left\{{\mathcal{\textit{z}}}(t)\right\}}^{\infty}_{t=0}$, where ${\mathcal{\textit{z}}}(t)\in \left\{0,1\right\}$:
	\begin{align}
		P\left[\mathcal{\textit{z}}(t)=z\right]=
		\begin{cases}
			p_t & z=1, \\
			1-p_t & z=0, \\
			0 & \text{otherwise.}
		\end{cases}
	\end{align}
	
	\textit{External request service proces}s is denoted by $\left\{\mathcal{\textit{r}}(t)\right\}_{t=0}^\infty$, where $\mathcal{\textit{r}}\left(t\right)\in\mathcal{\textit{R}}=\left\{r_0,r_1,r_2,...,r_K\right\}$ and $r_i$ indicates the service of the request from node $i\in\left\{1,2,...,K\right\}$, while $r_0$ stands for no request. We have:
	\begin{align}
		P\left[\mathcal{\textit{r}}(t)=r_i\right]=
		\begin{cases}
			q_i & i\in\left\{1,2,...,K\right\}, \\
			1-q & i=0, \\
			0 & \text{otherwise.}
		\end{cases}
	\end{align}
	
	\textit{Gossiping process} is denoted by $\left\{\mathcal{\textbf{\textit{g}}}(t)\right\}_{t=0}^\infty$, where $\mathcal{\textbf{\textit{g}}}(t)=\big(g_1(t),g_2(t),...,g_K(t)\big)\in\mathcal{G}$. The gossiping random variable $g_i(t)\in\left\{0,1\right\}$ is equal to $1$ if the node $i$ is updated by node $\left(i-1\right)$; it is equal to zero otherwise. In this case, $\mathcal{G}$ is the sample space of gossiping and is defined by:
	\begin{align}
		\mathcal{G}=\big\{\left(g_1,g_2,...,g_K\right):g_i\in\{0,1\}\big\}.
	\end{align}
	
	We also have the following:
	\begin{align} P\left[\mathcal{\textbf{\textit{g}}}=\left(g_1,g_2,...,g_K\right)\right]=\prod_{k=1}^{K}{\lambda_k^{g_k}\left(1-\lambda_k\right)^{1-g_k}}, \end{align}
	where we have assumed that the node $k$ is updated by its neighboring node $(k-1)$ with the probability $\lambda_k$. Observe that node $1$ is updated by node $K$.
	
	To formulate the transition probabilities, we use the total probability theorem as follows:
	\begin{align}
		P\left[s^\prime|s,a\right]=\sum_{r\in\mathcal{R}}\sum_{\substack{\mathbf{g} \in\mathcal{G} \\ e\in\left\{0,1\right\} \\ z\in\left\{0,1\right\}}} P\left[s^\prime|s,a,r,\mathbf{g},e,z\right]P_zP_eP_\mathbf{g}P_r,
	\end{align}
	where, for notation simplicity, we have used $P_z$, $P_e$, $P_\mathbf{g}$ and $P_r$ instead of $P\!\left[\mathcal{\textit{z}}(t)=z\right]$, $P\!\left[\mathcal{\textit{e}}(t)=e\right]$, $P\!\left[\mathcal{\textbf{\textit{g}}}(t)=\mathbf{g}\right]$, and $P\!\left[\mathcal{\textit{r}}(t)=r\right]$, respectively. We also drop the summation sets $\mathcal{G}$ and $\{0,1\}$.  
	\begin{align}
		\label{Trans_Overall_Eqn}
		P\left[s^\prime|s,a\right]=\sum_{r\in\mathcal{R}}\sum_{\ \mathbf{g},e,z\ } P\left[s^\prime|s,a,r,\mathbf{g},e,z\right]P_zP_eP_\mathbf{g}P_r,
	\end{align}
	where $s^\prime\!=\!\left(b^\prime,\Delta_1^\prime,...,\Delta_K^\prime,\Delta_C^\prime\right)$ and $s\!=\!\left(b,\Delta_1,...,\Delta_K,\Delta_C\right)$. We can expand (\ref{Trans_Overall_Eqn}) as follows: 
	\begin{align}
		\label{Trans_2terms_Eqn}
		P&\left[s^\prime|s,a\right]=\sum_{\ \mathbf{g},e,z\ }{P\left[s^\prime|s,a,r_0,\mathbf{g},e,z\right]P_zP_eP_\mathbf{g}P_{r_0}} \notag \\
		&+\sum_{r_i\in\mathcal{R}\setminus\{r_0\}}\sum_{\ \mathbf{g},e,z\ }{P\left[s^\prime|s,a,r_i,\mathbf{g},e,z\right]P_zP_eP_\mathbf{g}P_{r_i}}.
	\end{align}
	
	The first term in equation (\ref{Trans_2terms_Eqn}) stands for the transition probability when there is no external request. It can be simplified as follows:
	\begin{align}
		\label{Trans_FirsTerm_Eqn}
		P&\left[s^\prime\middle|s,a,r_0,\mathbf{g},e,z\right] \notag \\
		&=P\left[b^\prime,\Delta_1^\prime,...,\Delta_K^\prime,\Delta_C^\prime\middle| b,\Delta_1,...,\Delta_K,\Delta_C,a,r_0,\mathbf{g},e,z\right] \notag \\
		&=\underbrace{P\left[b^\prime\middle| b,r_0,e\right]}_{(E_1)}\times \underbrace{P\left[\Delta_C^\prime\middle| b,\Delta_C,r_0,z\right]}_{(E_2)} \notag \\
		&\times \underbrace{P\left[\Delta_1^\prime,\Delta_2^\prime,...,\Delta_K^\prime\middle|\Delta_1,\Delta_2,...,\Delta_K,r_0,\mathbf{g},z\right]}_{(E_3)}.
	\end{align}
	
	These probabilities are independent of action $a$, thus it has been omitted in the condition arguments. We have:
	\begin{align}
		(E_1)&=
		\begin{cases}
			1 & b^\prime=b+e \\
			0 & \text{otherwise,}
		\end{cases} \\
		(E_2)&=
		\begin{cases}
			1 & \Delta_C^\prime=\Delta_C+z \\
			0 & \text{otherwise,}
		\end{cases} \\
		(E_3)&=\prod_{k=1}^{K}P\left[\Delta_k^\prime|\Delta_k,\Delta_{k-1},g_k,z\right], 
	\end{align}
	where:
	\begin{align}
		P\left[\Delta_k^\prime|\Delta_k,\Delta_{k-1},g_k,z\right]=
		\begin{cases}
			1 & \Delta_k^\prime=\Delta_{g_k} \\
			0 & \text{otherwise.}
		\end{cases}
	\end{align}
	
	By substituting these transitions into the first term of equation (\ref{Trans_FirsTerm_Eqn}), we have the transition probabilities when there is no external request.
	\begin{align}
		\label{Trans_FirsTerm_Eqn1}
		P&\left[s^\prime\middle|\ s,a,r_0,\mathbf{g},e,z\right] \notag \\
		&=
		\begin{cases}
			1 & b^\prime=b+e,\ \Delta_C^\prime=\Delta_C+z,\ \Delta_k^\prime=\Delta_{g_k}, \\
			0 & \text{otherwise.}
		\end{cases}
	\end{align}
	
	We have defined $\left(\Delta_{g_1},\Delta_{g_2,},...,\Delta_{g_K}\right)$, where $\Delta_{g_k}=g_k\min{\left\{\Delta_k,\Delta_{k-1}\right\}}+\left(1-g_k\right)\Delta_k+z,\ k\in\{1,2,...,K\}$.
	
	Moreover, the second term in equation (\ref{Trans_2terms_Eqn}) stands for the transition probability when there is an external request. It can also be simplified as follows:
	\begin{align}
		P&\left[s^\prime|s,a,r_i,\mathbf{g},e,z\right] \\ &=P\left[b^\prime,\Delta_1^\prime,...,\Delta_K^\prime,\Delta_C^\prime|b,\Delta_1,...,\Delta_K,\Delta_C,a,r_i,\mathbf{g},e,z\right] \notag \\
		&=\underbrace{P\left[b^\prime|b,a,r_i,e\right]}_{(E_4)}\times \underbrace{P\left[\Delta_i^\prime,\Delta_C^\prime|b,\Delta_i,\Delta_C,a,r_i,z\right]}_{(E_5)} \notag \\ 
		&\times \underbrace{P\left[\Delta_1^\prime,...,\Delta_{i-1}^\prime,\Delta_{i+1}^\prime,...,\Delta_K^\prime|\Delta_1,...,\Delta_K,r_i,\mathbf{g},z\right]}_{(E_6)}. \notag
	\end{align}
	
	When $a=0$, then we have:
	\begin{align}
		(E_4)&=
		\begin{cases}
			1 & b^\prime=b+e, \\
			0 & \text{otherwise,}
		\end{cases}\\
		(E_5)&=
		\begin{cases}
			1 & \parbox[][][t]{4.4cm}{$\Delta_C^\prime\!=\!\Delta_C\!+\!z,\\ \Delta_i^\prime\!=\!\min\!{\left\{\Delta_i,\Delta_C\right\}}\!+\!z\!=\!\Delta_C^\prime,$} \\
			0 & \text{otherwise,}
		\end{cases} \\ 
		(E_6)&=\prod_{\substack{k=1 \\ k\neq i}}^{K}P\left[\Delta_k^\prime|\Delta_k,\Delta_{k\!-\!1},g_k,z\right],
    \end{align}
    where:
    \begin{align}
		P\left[\Delta_k^\prime|\Delta_k,\Delta_{k-1},g_k,z\right]\!&=\!
		\begin{cases}
			1 & \Delta_k^\prime=\Delta_{g_k}, \\
			0 & \text{otherwise.}
		\end{cases}
	\end{align}
	
	In the case of $a=1$ and $b=0$ (empty battery), the transition probabilities are similar to the case $a=0$. If $a=1$ and $b\geq 1$ (non-empty battery), then we have:
	\begin{align}
		(E_4)&=
		\begin{cases}
			1 & b^\prime=b-1+e, \\
			0 & \text{otherwise,}
		\end{cases} \\
		(E_5)&=
		\begin{cases}
			1 & \Delta_i^\prime=\Delta_C^\prime=z, \\
			0 & \text{otherwise,}
		\end{cases} \\
		(E_6)&=\prod_{\substack{k=1 \\ k\neq i}}^{K}P\left[\Delta_k^\prime|\Delta_k,\Delta_{k-1},g_k,z\right],
    \end{align}
    where:
    \begin{align}
		P\left[\Delta_k^\prime|\Delta_k,\Delta_{k-1},g_k,z\right]&=
		\begin{cases}
			1 & \Delta_k^\prime=\Delta_{g_k}, \\
			0 & \text{otherwise.}
		\end{cases}
	\end{align}
	
	\noindent To summarize, if $a=0$ or $b=0$, then we have:
	\begin{align}
		\label{Trans_SecondTerm_a0b0_Eqn}
		P&\left[s^\prime|s,a,r_i,\mathbf{g},e,z\right] \\
		&=
		\begin{cases}
			1 & b^\prime=b+e,\ \Delta_i^\prime=\Delta_C^\prime=\Delta_C+z,\ \Delta_k^\prime=\Delta_{g_k}, \\
			0 & \text{otherwise,}
		\end{cases} \notag
	\end{align}
	
	\noindent and in the case that $a=1$ and $b\geq1$, then:
	\begin{align}
		\label{Trans_SecondTerm_a1b1_Eqn}
		P&\left[s^\prime|s,a,r_i,\mathbf{g},e,z\right] \\
		&=
		\begin{cases}
			1 & b^\prime=b-1+e,\ \Delta_i^\prime=\Delta_C^\prime=z,\ \Delta_k^\prime=\Delta_{g_k}, \\
			0 & \text{otherwise,} \notag
		\end{cases}
	\end{align}
	
	\noindent where in equations (\ref{Trans_SecondTerm_a0b0_Eqn}) and (\ref{Trans_SecondTerm_a1b1_Eqn}), $k\in\left\{1,2,...,i-1,i+1,...,K\right\}$.
	By substituting (\ref{Trans_FirsTerm_Eqn1}), (\ref{Trans_SecondTerm_a0b0_Eqn}), and (\ref{Trans_SecondTerm_a1b1_Eqn}) into (\ref{Trans_2terms_Eqn}), we obtain the transition probabilities of the MDP problem.

	\section{Analytical Results}
	\label{AnalyticalResults}
	In this section, we study the properties of the value function and the optimal policy for the problem (\ref{MDP_eqn}).
	
	\textbf{Definition.} \textit{An MDP is weakly accessible (or weakly communicating) if the set of states can be partitioned into two subsets $S_t$ and $S_c$ such that: (a) all states in $S_t$ are transient under every stationary policy, and (b) for every two states $s$ and $s^\prime$ in $S_c$, $s^\prime$ is accessible from $s$. We say that state $s^\prime$ is accessible from state $s$, if $s^\prime$ can be reached from $s$ under some stationary policy.}
	
	\textbf{Proposition 1.} \textit{The MDP problem (\ref{MDP_eqn}) is weakly accessible.}
	
	\textit{Proof:} The proof can be found in appendix \ref{Appen0}. \hfill $\blacksquare$\\
	
	\textbf{Proposition 2.} \textit{In the MDP problem (\ref{MDP_eqn}), the optimal average cost $J^\ast$ achieved by an optimal policy $\pi^\ast$ is the same for all initial states, and it satisfies the Bellman’s equation:}
	\begin{equation}
		\label{Bellman_eqn}
		J^\ast\!+\!V(s)\!=\!\min_{a\in\left\{0,1\right\}}{\!\left\{\!\Delta_{AVG}(s)\!+\!\sum_{s^\prime\in S}{P\left[s^\prime \big | s,a\right]\!V(s^\prime)}\!\right\}},
	\end{equation}
	
	\noindent \textit{where $V(s)$ is the value function of the MDP problem. The optimal policy $\pi^\ast(s)$ can also be obtained by solving the Bellman’s equation:}
	\begin{equation}
		\label{OptimalAction_eqn}
		\pi^\ast(s)\! \in \! \argmin_{\!a\in\left\{0,1\right\}}{\left\{\!\Delta_{AVG}(s)\!+\!\sum_{s^\prime\in S}{P\left[s^\prime \big | s,a\right]\!V(s^\prime)}\right\}}.
	\end{equation}%
	
	\textit{Proof:} According to Proposition 1, the problem (\ref{MDP_eqn}) is weakly accessible. Therefore, by Prop. 4.2.3 in \cite{bertsekas2011dynamic}, the optimal average cost is the same for all initial states. In addition, according to Prop. 4.2.6 in \cite{bertsekas2011dynamic}, an optimal policy exists, and by Prop. 4.2.1 in \cite{bertsekas2011dynamic}, if we can find such $J^\ast$ and $V(s)$ that satisfy (\ref{Bellman_eqn}), then the optimal policy is given by (\ref{OptimalAction_eqn}). \hfill $\blacksquare$\\
	
	It can be seen that the optimal policy $\pi^\ast$ depends on $V(s)$, which generally does not have a closed-form solution. Standard methods such as the Value Iteration and Policy Iteration algorithms can be used to solve this optimization problem. We also provide RVI algorithm \ref{alg_VIA} for integrity. In this algorithm, $ sp(V_t-V_{t-1})<\epsilon $ is the loop termination condition, where $sp\left(V(s)\right)\triangleq\max_{s\in S}{V(s)}-\min_{s\in S}{V(s)}$.
	
	{\small 
		\begin{algorithm}[!t]
			\caption{Relative VI algorithm}
			\label{alg_VIA}
			\begin{algorithmic}
				\STATE set $ v_0(s)=0\ \forall s \in S $
				\STATE set $ t=0, \epsilon>0 $
				\STATE set $ V_0(s)=0\ \forall s \in S $
				\REPEAT
				\STATE{
					$ t\leftarrow{t+1} $
					
					\FORALL{$ s \in S $}
					\STATE{
						$
						v_t(s)=\min_{a\in\left\{0,1\right\}}{\sum_{s^\prime\in S}{P\left(s^\prime|s,a\right)\left[C(s,a)+V(s^\prime)\right]}}
						$
						
						$
						V_t(s)=v_t(s)-v_t(s_0)
						$
						
						where $ s_0 $ is a fixed state chosen arbitrarily
					}
					\ENDFOR
					
				}
				\UNTIL{$ sp(V_t-V_{t-1})<\epsilon $}
				\RETURN $\arg \min V(s) $
			\end{algorithmic}
		\end{algorithm}
	}
	
	\textbf{Definition.} \textit{The state $s=\left(b,\Delta_1,\Delta_2,\ldots,\Delta_K,\Delta_C\right)$ is a causal state if $\Delta_i\geq\Delta_C$ for all $i\in\{1,2,\ldots,K\}$}.
	
	\textbf{Theorem 1.} \textit{For a causal state $s=\left(b,\Delta_1,\Delta_2,\ldots,\Delta_K,\Delta_C\right)$, the optimal policy $\pi^\ast(s)$ is not a function of $(\Delta_1,\Delta_2,\ldots,\Delta_K)$ values, i.e., the optimal actions are only a function of $b$ and $\Delta_C$ (for fixed $p_t$, $\beta$, $\lambda_k$ and $q_k$ parameters).}
	
	\textit{Proof:} The proof can be found in appendix \ref{Appen1}. \hfill $\blacksquare$\\
	
	\textbf{Definition.} \textit{Suppose that there exists a $\Delta_{C_{Thr}}(b)$ for each battery state $b$ such that the action $a^\pi(b,\Delta_C)$ is $1$ for $\Delta_C \geq \Delta_{C_{Thr}}(b)$, and $0$ otherwise. In this case, the policy $\pi$ is a threshold policy.}
	
	\textbf{Lemma 1.} Value function is increasing in $\Delta_C$, i.e., if $\Delta_{C1}\le\Delta_{C2}$, then: 
	\begin{align}
		V&\left(b,\Delta_1,\ldots,\Delta_{i-1},\Delta_{C1},\Delta_{i+1},\ldots,\Delta_K,\Delta_{C1}\right) \notag\\
		&\le V\left(b,\Delta_1,\ldots,\Delta_{i-1},\Delta_{C2},\Delta_{i+1},\ldots,\Delta_K,\Delta_{C2}\right).
	\end{align}%
	
	\textit{Proof:} The proof can be found in appendix \ref{pLemma1_Appen}. \hfill $\blacksquare$\\
	
	\textbf{Theorem 2.} \textit{The optimal policy at the aggregator is a threshold policy.}
	
	\textit{Proof:} The proof can be found in appendix \ref{Appen2}. \hfill $\blacksquare$
	
	\section{Numerical Results}
	\label{NumResults}
	
	In this section, we investigate the characteristics of the optimal policy and highlight the effect of the system parameters on the average Version AoI of the network. To this end, we provide numerical results to verify that the optimal policy has a threshold-based structure. Then, we investigate the effect of the external requests' service probability ($q_i, i \in \{1,2,..., K\}$), the energy arrival probability ($\beta$), the battery size of the sensor ($B$), the gossiping parameters ($\lambda_i, i \in \{1,2,..., K\}$), and the source state change probability ($p_t$) on the optimal average Version AoI. We also compare the performance of our optimal policy with two baseline policies: greedy and random. In the greedy policy, the aggregator C requests a fresh update whenever an external request arrives, regardless of its Version AoI state. In the random policy, C requests a fresh update with a probability of $0.5$. In both cases, after receiving a request from C, the sensor S samples and sends a new update only if the battery is not empty. We run our algorithms over $4000$ time slots and average over $400$ simulations. Also, the value of $\Delta_{max}$ has been set to $9$.
	
	\subsection{Structure of the Optimal Update Policy}
\begin{figure}
		\centering	
		\subfloat[$p_t=0.5,\beta=0.2$]{\label{Action_fig1.eps}\includegraphics[width=2.6in]{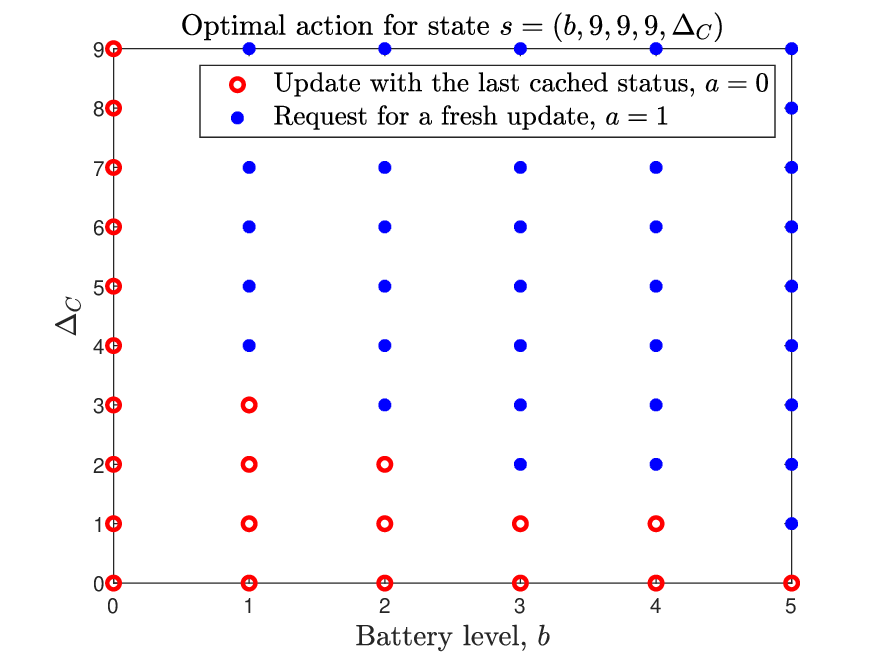}}\\
		\subfloat[$p_t=0.5,\beta=0.1$]{\label{Action_fig2}\includegraphics[width=2.6in]{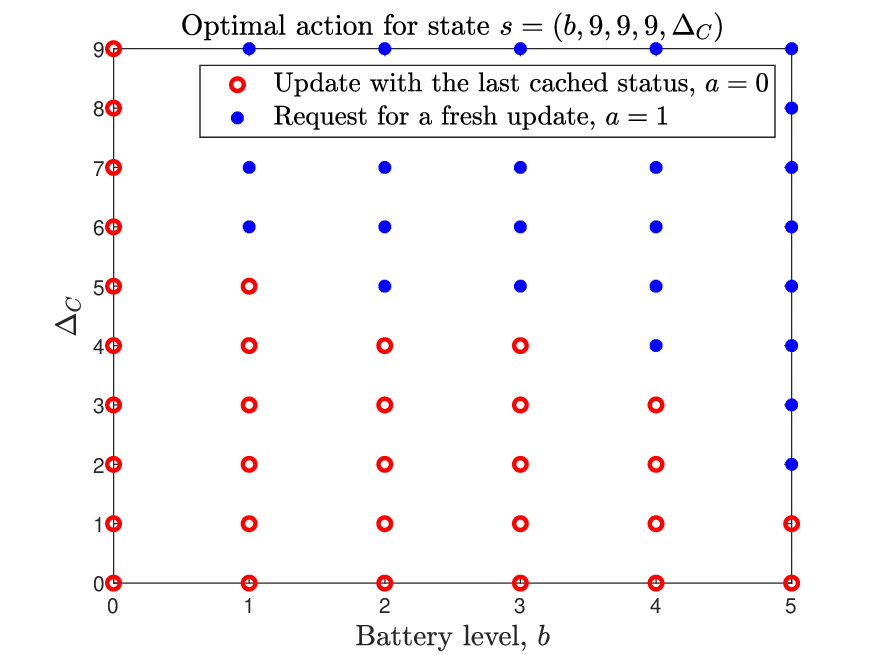}}\\
		\caption{The structure of the optimal policy.}
		\label{Action_fig}
	\end{figure}

	We provide numerical results to analyze the structure of the optimal policy. We utilize the Value Iteration algorithm to solve the MDP problem and find the optimal policy. According to Theorem 1, the optimal policy does not depend on $\Delta_i$ values, as far as $\Delta_i \geq \Delta_C$, for $i \in \{1,2,..., K\}$. This has also been verified via simulations with different values for system parameters. In addition, according to Theorem 2, the optimal policy has a threshold-based structure, which is confirmed by simulation results. Fig. \ref{Action_fig1} depicts the optimal actions for a network with $K=3$ nodes. The system parameters are set to $\beta=0.2$, $p_t=0.5$, $q_1=0.1$, $q_2=0.2$, $q_3=0.3$, and $B=5$. Each battery state has a threshold value for Version AoI at the aggregator where the action changes from $0$ to $1$. In Fig. \ref{Action_fig2}, the value of $\beta$ is reduced to $0.1$. Reducing $\beta$ imposes a severe energy limitation on the system, and optimizing the average Version AoI is achieved by choosing higher thresholds.
	
	\subsection{The Impact of $\left\{q_i\right\}_{i=1}^K$ on the Average Version AoI}
	
\begin{figure}[!t]
		\centering
		\includegraphics[width=2.6in]{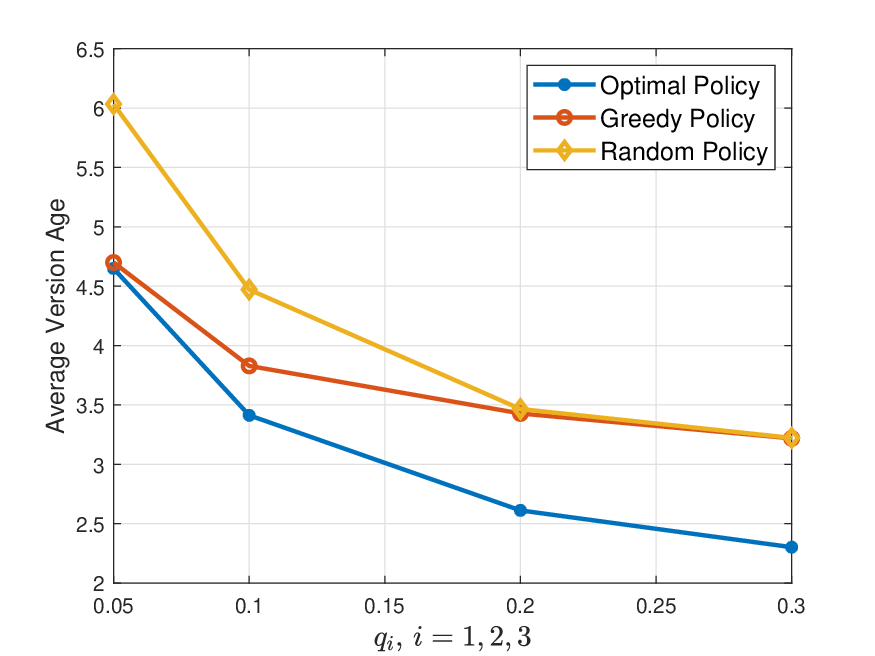}
		\caption{Average Version AoI vs. $ q_i $.}
		\label{VersionAge_q}
	\end{figure}

\begin{figure}[!t]
		\centering
		\includegraphics[width=3.4in]{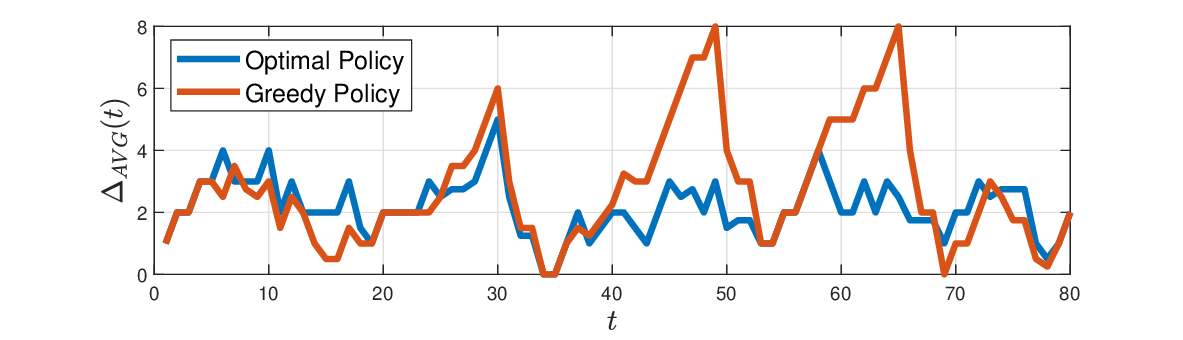}
		\caption{A sample path of Version AoI.}
		\label{Sample_VersionAge}
		
	\end{figure}

	In order to investigate the effect of $\left\{q_i\right\}_{i=1}^K$ on the Version AoI of the system, we consider a network with $K=3$ nodes where $q_1=q_2=q_3$. We fix the parameters $\beta$, $p_t$, and $B$ to $0.2$, $0.5$ and $5$, respectively. The gossiping parameters $\lambda_i$ are also fixed to $0.2$ for $i=1, 2, 3$. As shown in Fig. \ref{VersionAge_q}, the optimal policy results in a lower average Version AoI than greedy and random policies. The following observations can be drawn from the figure:
	\begin{itemize}
		\item The increase of $q_i$ values improves the average Version AoI. However, for exceedingly high $q_i$, the average Version AoI will be restricted by the energy budget so that it will remain constant.
		\item When the service probability of the external request is low compared to $\beta$, the battery has a high probability of not being empty when an external request arrives. In this case, responding to the \textit{rare} external requests with an immediate update is expected. This is why the performance of the greedy and optimal policies is the same. Obviously, for this case, the random policy is far from optimal because if an external request is not served by a new update while the battery is not empty, it may take too long to receive the subsequent external request, and the Version AoI increases during this interval.
		\item When the service probability of external requests increases, selecting the best policy for energy management becomes more crucial. If external requests occur \textit{more frequently} compared to energy arrivals, serving a request with an immediate update may not be an effective policy. This is because the update could happen at low Version AoI levels, leading to a long period of battery depletion without any updates. As a result, Version AoI can increase to a high value during this period. Therefore, using a higher Version AoI threshold to update the network can be a more effective strategy, which is the optimal policy. As an illustration, Fig. \ref{Sample_VersionAge} shows the time evolution of a sample average Version AoI function for both optimal and greedy policies.	
	\end{itemize}
	

	\subsection{The Impact of $B$ on the Average Version AoI}
	
\begin{figure}[!t]
		\centering
		\includegraphics[width=2.6in]{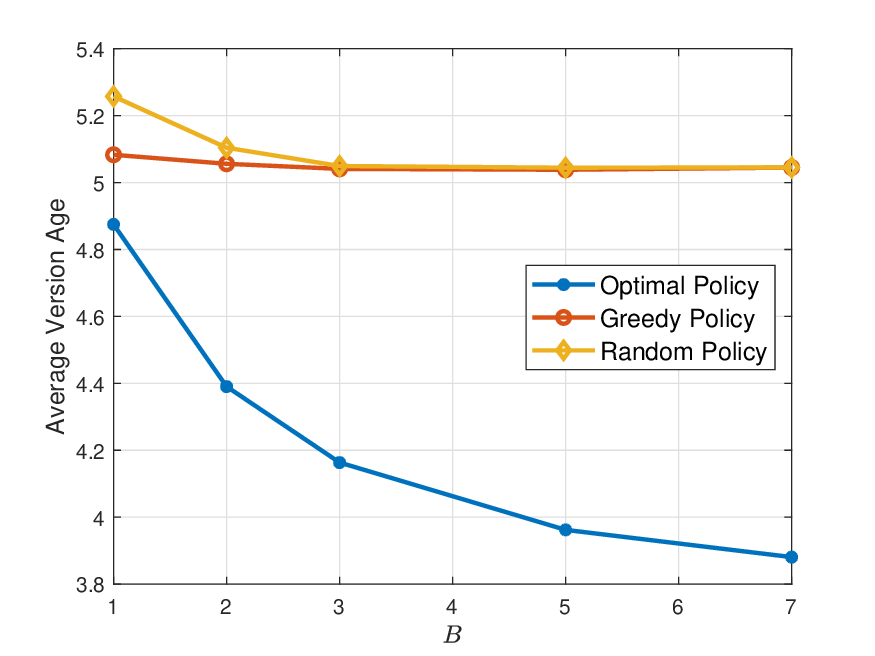}
		\caption{Average Version AoI vs. $ B $.}
		\label{VersionAge_B}
	\end{figure}

	The effect of the battery size $ B $ on the average Version AoI is investigated for $\beta=0.1$, $p_t=0.5$, $q_1=0.1$, $q_2=0.2$, $q_3=0.3$, and $\lambda_i=0.2$ in a network with $K=3$ nodes. As expected, in Fig. \ref{VersionAge_B}, we see that the optimal policy outperforms the greedy and random policies, and the average Version AoI decreases as $ B $ increases. It can also be observed that for a larger battery size, the gap of average Version AoI between the optimal policy and the other policies becomes larger. The reason is that aggregator C can optimize the different Version AoI thresholds for different battery levels without being limited to the capacity of the sensor's battery. However, depending on the balance between $ \beta $ and $ q_i $ values, the harvested energy rarely exceeds a specific level. Therefore, a large battery size does not impact the system's performance. This is why the average Version AoI in Fig. \ref{VersionAge_B} asymptotically converges to a constant value.
	
	\subsection{The Impact of $\beta$ on the Average Version AoI}
	
\begin{figure}[!t]
		\centering
		\includegraphics[width=2.6in]{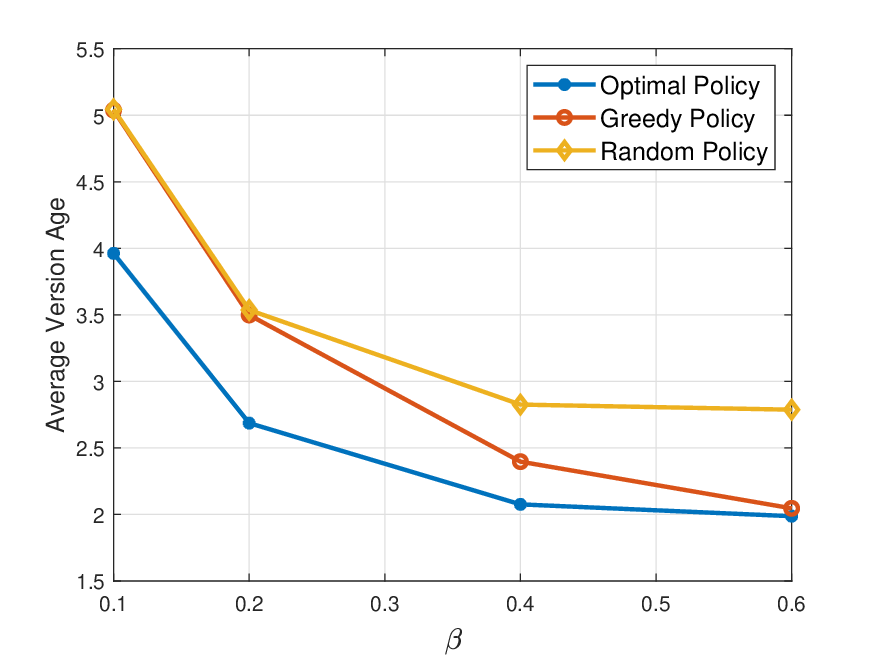}
		\caption{Average Version AoI vs. $ \beta $.}
		\label{VersionAge_beta}
		
	\end{figure}

	We analyze the effect of the parameter $\beta$ on the average Version AoI for a network with $K=3$ nodes and $p_t=0.5$, $q_1=0.1$, $q_2=0.2$, $q_3=0.3$, $B=5$, and $\lambda_{i=1,2,3}=0.2$, which is depicted in Fig. \ref{VersionAge_beta}. When $\beta$ is large enough compared to $q_i$, there is less concern about the energy limitation, and therefore a greedy policy becomes optimal. This is shown in Fig. \ref{VersionAge_beta}, where the optimal and greedy policies converge to the optimal average Version AoI for larger values of $ \beta $. On the other hand, when $ \beta $ is low, energy management becomes critical to maintain a reasonable value for the average Version AoI. In this case, the optimal policy outperforms the greedy and random policies with a considerable gap.
	

	\subsection{The Impact of $p_t$ on the Average Version AoI}
	
\begin{figure}[!t]
		\centering
		\includegraphics[width=2.6in]{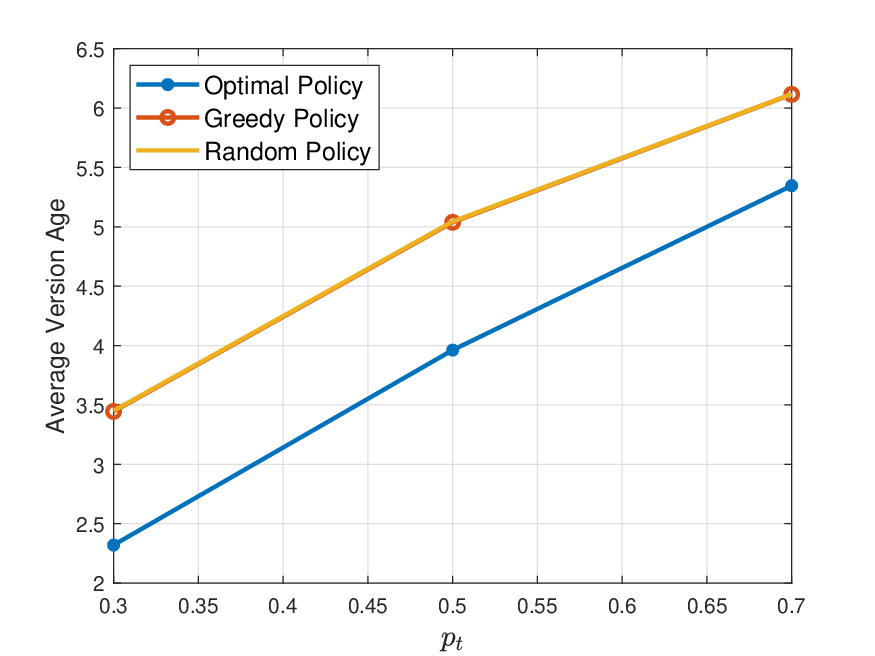}
		\caption{Average Version AoI vs. $ p_t $.}
		\label{VersionAge_pt}
	\end{figure}

	The probability of state change at the source, denoted by $p_t$, is another parameter that affects the average Version age of the system. In Fig. \ref{VersionAge_pt}, the average Version AoI in a $3$-node network for different $p_t$ values is plotted. We consider $\beta=0.1$, $q_1=0.1$, $q_2=0.2$, $q_3=0.3$, $B=5$, and $\lambda_{i=1,2,3}=0.2$ for this simulation. The average Version AoI increases with $p_t$. This is because the Version age at the destination nodes increases with a higher probability at each time slot while the energy resource remains unchanged, and the aggregator will not be able to update the nodes at a higher rate considering the state of the sensor's battery.
	
	\subsection{The Impact of $\left\{\lambda_i\right\}_{i=1}^K$ on the Average Version AoI}
	
\begin{figure}[!t]
		\centering
		\includegraphics[width=2.6in]{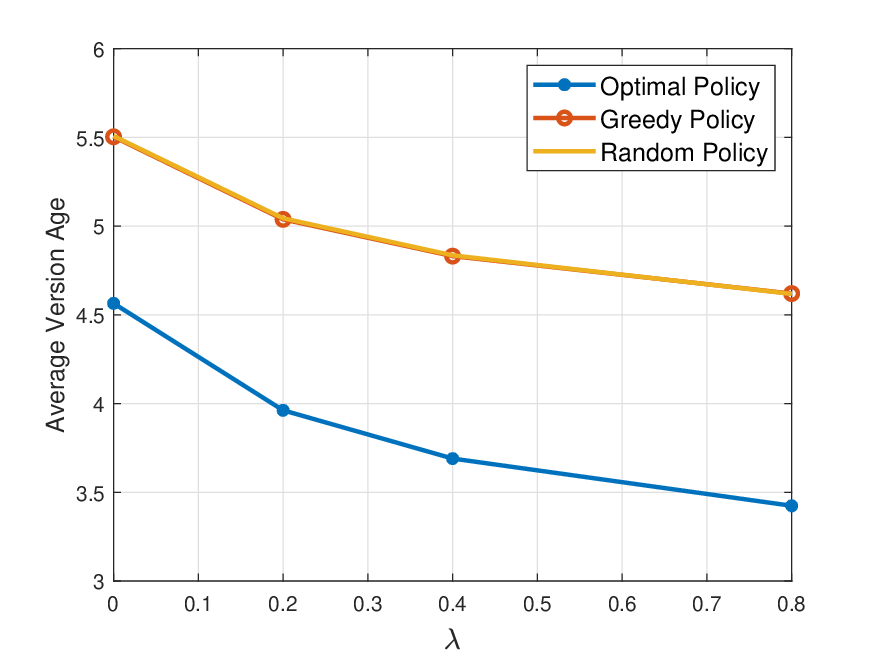}
		\caption{Average Version AoI vs. $ \lambda_i $.}
		\label{VersionAge_lambda}
		
	\end{figure}

	Gossiping can improve the performance regarding Version age at the destination nodes and the average Version AoI in the system. It has been illustrated in Fig. \ref{VersionAge_lambda} where the average Version AoI in a $3$-node network for different $\lambda_i$ is depicted. We have $\lambda_1=\lambda_2=\lambda_3=\lambda$ and the system parameters are fixed to $\beta=0.1$, $p_t=0.5$, $q_1=0.1$, $q_2=0.2$, $q_3=0.3$, and $B=5$.
	

	\section{Conclusion}
	\label{conclusion}
	In this work, we considered a real-time monitoring system where an aggregator aims to minimize the average Version AoI in a gossiping network. The aggregator serves the requests of nodes within the network with either a cached or a fresh status update captured by an energy harvesting sensor. We utilized the Markov Decision Process framework to model the problem and determine the optimal policy at the aggregator. We have demonstrated that the Version age at the destination nodes does not affect the optimal policy and that the optimal policy is a threshold policy based on the Version AoI at the aggregator. Finally, we have numerically investigated the effect of system parameters on the average Version AoI and the performance of the optimal policy.
	
	\bibliographystyle{IEEEtran}
	\bibliography{Refs}

\begin{thebibliography}{10}
\providecommand{\url}[1]{#1}
\csname url@samestyle\endcsname
\providecommand{\newblock}{\relax}
\providecommand{\bibinfo}[2]{#2}
\providecommand{\BIBentrySTDinterwordspacing}{\spaceskip=0pt\relax}
\providecommand{\BIBentryALTinterwordstretchfactor}{4}
\providecommand{\BIBentryALTinterwordspacing}{\spaceskip=\fontdimen2\font plus
\BIBentryALTinterwordstretchfactor\fontdimen3\font minus
  \fontdimen4\font\relax}
\providecommand{\BIBforeignlanguage}[2]{{%
\expandafter\ifx\csname l@#1\endcsname\relax
\typeout{** WARNING: IEEEtran.bst: No hyphenation pattern has been}%
\typeout{** loaded for the language `#1'. Using the pattern for}%
\typeout{** the default language instead.}%
\else
\language=\csname l@#1\endcsname
\fi
#2}}
\providecommand{\BIBdecl}{\relax}
\BIBdecl

\bibitem{WiOpt23}
E.~Delfani and N.~Pappas, ``Version age-optimal cached status updates in a
  gossiping network with energy harvesting sensor,'' in \emph{21st
  International Symposium on Modeling and Optimization in Mobile, Ad hoc, and
  Wireless Networks (WiOpt)}, 2023.

\bibitem{abd2019role}
M.~A. Abd-Elmagid, N.~Pappas, and H.~S. Dhillon, ``On the role of age of
  information in the internet of things,'' \emph{IEEE Communications Magazine},
  vol.~57, no.~12, pp. 72--77, 2019.

\bibitem{shreedhar2019age}
T.~Shreedhar, S.~K. Kaul, and R.~D. Yates, ``An age control transport protocol
  for delivering fresh updates in the internet-of-things,'' in \emph{IEEE
  International Symposium on "A World of Wireless, Mobile and Multimedia
  Networks"(WoWMoM)}, 2019.

\bibitem{zhang2015mission}
Q.~Zhang and F.~H. Fitzek, ``Mission critical iot communication in 5g,'' in
  \emph{Future Access Enablers of Ubiquitous and Intelligent
  Infrastructures}.\hskip 1em plus 0.5em minus 0.4em\relax Springer, 2015, pp.
  35--41.

\bibitem{kaul2012real}
S.~Kaul, R.~Yates, and M.~Gruteser, ``Real-time status: How often should one
  update?'' in \emph{IEEE INFOCOM}, 2012, pp. 2731--2735.

\bibitem{pappas2023age}
N.~Pappas, M.~A. Abd-Elmagid, B.~Zhou, W.~Saad, and H.~S. Dhillon, \emph{Age of
  Information: Foundations and Applications}.\hskip 1em plus 0.5em minus
  0.4em\relax Cambridge University Press, 2023.

\bibitem{maatouk2020age}
A.~Maatouk, S.~Kriouile, M.~Assaad, and A.~Ephremides, ``The age of incorrect
  information: A new performance metric for status updates,'' \emph{IEEE/ACM
  Transactions on Networking}, vol.~28, no.~5, pp. 2215--2228, 2020.

\bibitem{zhong2018two}
J.~Zhong, R.~D. Yates, and E.~Soljanin, ``Two freshness metrics for local cache
  refresh,'' in \emph{IEEE International Symposium on Information Theory
  (ISIT)}, 2018.

\bibitem{liu20222}
Q.~Liu, C.~Li, Y.~T. Hou, W.~Lou, J.~H. Reed, and S.~Kompella, ``Ao2i:
  Minimizing age of outdated information to improve freshness in data
  collection,'' in \emph{IEEE Conference on Computer Communications}, 2022.

\bibitem{stamatakis2019control}
G.~Stamatakis, N.~Pappas, and A.~Traganitis, ``Control of status updates for
  energy harvesting devices that monitor processes with alarms,'' in \emph{IEEE
  Globecom Workshops}, 2019.

\bibitem{yates2021age}
R.~D. Yates, ``The age of gossip in networks,'' in \emph{IEEE International
  Symposium on Information Theory (ISIT)}, 2021.

\bibitem{yates2021timely}
------, ``Timely gossip,'' in \emph{IEEE 22nd International Workshop on Signal
  Processing Advances in Wireless Communications (SPAWC)}, 2021.

\bibitem{buyukates2022version}
B.~Buyukates, M.~Bastopcu, and S.~Ulukus, ``Version age of information in
  clustered gossip networks,'' \emph{IEEE Journal on Selected Areas in
  Information Theory}, vol.~3, no.~1, 2022.

\bibitem{kaswan2022timely}
P.~Kaswan and S.~Ulukus, ``Timely gossiping with file slicing and network
  coding,'' in \emph{IEEE International Symposium on Information Theory
  (ISIT)}, 2022.

\bibitem{kaswan2022susceptibility}
------, ``Susceptibility of age of gossip to timestomping,'' in \emph{IEEE
  Information Theory Workshop (ITW)}, 2022, pp. 398--403.

\bibitem{mitra2022asuman}
P.~Mitra and S.~Ulukus, ``Asuman: Age sense updating multiple access in
  networks,'' in \emph{58th Annual Allerton Conference on Communication,
  Control, and Computing}, 2022.

\bibitem{kaswan10051981}
P.~Kaswan and S.~Ulukus, ``Age of gossip in ring networks in the presence of
  jamming attacks,'' in \emph{56th Asilomar Conference on Signals, Systems, and
  Computers}, 2022, pp. 1055--1059.

\bibitem{mitra2023age}
P.~Mitra and S.~Ulukus, ``Age-aware gossiping in network topologies,''
  \emph{arXiv preprint arXiv:2304.03249}, 2023.

\bibitem{mitra2023learning}
------, ``A learning based scheme for fair timeliness in sparse gossip
  networks,'' \emph{arXiv preprint arXiv:2310.01396}, 2023.

\bibitem{abd2023distribution}
M.~A. Abd-Elmagid and H.~S. Dhillon, ``Distribution of the age of gossip in
  networks,'' \emph{Entropy}, vol.~25, no.~2, p. 364, 2023.

\bibitem{gunduz2014designing}
D.~Gunduz, K.~Stamatiou, N.~Michelusi, and M.~Zorzi, ``Designing intelligent
  energy harvesting communication systems,'' \emph{IEEE communications
  magazine}, vol.~52, no.~1, pp. 210--216, 2014.

\bibitem{bacinoglu2015age}
B.~T. Bacinoglu, E.~T. Ceran, and E.~Uysal-Biyikoglu, ``Age of information
  under energy replenishment constraints,'' in \emph{Information Theory and
  Applications Workshop (ITA)}, 2015.

\bibitem{yates2015lazy}
R.~D. Yates, ``Lazy is timely: Status updates by an energy harvesting source,''
  in \emph{IEEE International Symposium on Information Theory (ISIT)}, 2015.

\bibitem{bacinoglu2017scheduling}
B.~T. Bacinoglu and E.~Uysal-Biyikoglu, ``Scheduling status updates to minimize
  age of information with an energy harvesting sensor,'' in \emph{IEEE
  international symposium on information theory (ISIT)}, 2017.

\bibitem{sun2017update}
Y.~Sun, E.~Uysal-Biyikoglu, R.~D. Yates, C.~E. Koksal, and N.~B. Shroff,
  ``Update or wait: How to keep your data fresh,'' \emph{IEEE Transactions on
  Information Theory}, vol.~63, no.~11, pp. 7492--7508, 2017.

\bibitem{bacinoglu2018achieving}
B.~T. Bacinoglu, Y.~Sun, E.~Uysal-Bivikoglu, and V.~Mutlu, ``Achieving the
  age-energy tradeoff with a finite-battery energy harvesting source,'' in
  \emph{IEEE International Symposium on Information Theory (ISIT)}, 2018.

\bibitem{arafa2018age}
A.~Arafa, J.~Yang, and S.~Ulukus, ``Age-minimal online policies for energy
  harvesting sensors with random battery recharges,'' in \emph{IEEE
  international conference on communications (ICC)}, 2018.

\bibitem{farazi2018average}
S.~Farazi, A.~G. Klein, and D.~R. Brown, ``Average age of information for
  status update systems with an energy harvesting server,'' in \emph{IEEE
  INFOCOM Workshops}, 2018.

\bibitem{feng2018minimizing}
S.~Feng and J.~Yang, ``Minimizing age of information for an energy harvesting
  source with updating failures,'' in \emph{IEEE International Symposium on
  Information Theory (ISIT)}, 2018.

\bibitem{chen2019age}
Z.~Chen, N.~Pappas, E.~Bj{\"o}rnson, and E.~G. Larsson, ``Age of information in
  a multiple access channel with heterogeneous traffic and an energy harvesting
  node,'' in \emph{IEEE INFOCOM Workshops}, 2019.

\bibitem{feng2021age}
S.~Feng and J.~Yang, ``Age of information minimization for an energy harvesting
  source with updating erasures: Without and with feedback,'' \emph{IEEE
  Transactions on Communications}, vol.~69, no.~8, pp. 5091--5105, 2021.

\bibitem{arafa2019age}
A.~Arafa, J.~Yang, S.~Ulukus, and H.~V. Poor, ``Age-minimal transmission for
  energy harvesting sensors with finite batteries: Online policies,''
  \emph{IEEE Transactions on Information Theory}, vol.~66, no.~1, pp. 534--556,
  2019.

\bibitem{bacinoglu2019optimal}
B.~T. Bacinoglu, Y.~Sun, E.~Uysal, and V.~Mutlu, ``Optimal status updating with
  a finite-battery energy harvesting source,'' \emph{Journal of Communications
  and Networks}, vol.~21, no.~3, pp. 280--294, 2019.

\bibitem{abd2019online}
M.~A. Abd-Elmagid, H.~S. Dhillon, and N.~Pappas, ``Online age-minimal sampling
  policy for rf-powered iot networks,'' in \emph{IEEE Global Communications
  Conference (GLOBECOM)}, 2019.

\bibitem{abd2020reinforcement}
------, ``A reinforcement learning framework for optimizing age of information
  in rf-powered communication systems,'' \emph{IEEE Transactions on
  Communications}, vol.~68, no.~8, 2020.

\bibitem{leng2019minimizing}
S.~Leng and A.~Yener, ``Minimizing age of information for an energy harvesting
  cognitive radio,'' in \emph{IEEE Wireless Communications and Networking
  Conference (WCNC)}, 2019.

\bibitem{ceran2019reinforcement}
E.~T. Ceran, D.~G{\"u}nd{\"u}z, and A.~Gy{\"o}rgy, ``Reinforcement learning to
  minimize age of information with an energy harvesting sensor with harq and
  sensing cost,'' in \emph{IEEE INFOCOM Workshops}, 2019, pp. 656--661.

\bibitem{gindullina2021age}
E.~Gindullina, L.~Badia, and D.~G{\"u}nd{\"u}z, ``Age-of-information with
  information source diversity in an energy harvesting system,'' \emph{IEEE
  Transactions on Green Communications and Networking}, vol.~5, no.~3, pp.
  1529--1540, 2021.

\bibitem{hatami2022demand}
M.~Hatami, M.~Leinonen, Z.~Chen, N.~Pappas, and M.~Codreanu, ``On-demand aoi
  minimization in resource-constrained cache-enabled iot networks with energy
  harvesting sensors,'' \emph{IEEE Transactions on Communications}, vol.~70,
  no.~11, pp. 7446--7463, 2022.

\bibitem{hatami2021aoi}
M.~Hatami, M.~Leinonen, and M.~Codreanu, ``Aoi minimization in status update
  control with energy harvesting sensors,'' \emph{IEEE Transactions on
  Communications}, vol.~69, no.~12, pp. 8335--8351, 2021.

\bibitem{holm2021freshness}
J.~Holm, A.~E. Kal{\o}r, F.~Chiariotti, B.~Soret, S.~K. Jensen, T.~B. Pedersen,
  and P.~Popovski, ``Freshness on demand: Optimizing age of information for the
  query process,'' in \emph{IEEE International Conference on Communications},
  2021.

\bibitem{bertsekas2011dynamic}
D.~P. Bertsekas, \emph{Dynamic Programming and Optimal Control, Vol. II},
  3rd~ed.\hskip 1em plus 0.5em minus 0.4em\relax Athena Scientific, 2007.

\end{thebibliography}
	
	\appendices
	
	\section{Proof of Proposition 1}
	\label{Appen0}
	
	The network nodes are updated via the aggregator C, which means that they have the same version or older versions compared to C. Notice that at least one node has the same Version AoI as C. Thus, any state $s=\left(b,\Delta_1,\Delta_2,\ldots,\Delta_K,\Delta_C\right)$ is a transient state if $\Delta_k<\Delta_C$ for some $k\in\{1,2,\ldots, K\}$, or $\Delta_k>\Delta_C$ for all $k\in\{1,2,\ldots, K\}$. In what follows, we show that any non-transient state $s^\prime=\left(b^\prime,\Delta_1^\prime,\Delta_2^\prime,\ldots,\Delta_K^\prime,\Delta_C^\prime\right)$ is accessible from any other non-transient state $s=\left(b,\Delta_1,\Delta_2,\ldots,\Delta_K,\Delta_C\right)$ under a stationary stochastic policy $\pi$ where the action $a\in\{0,1\}$ at each state is randomly selected with probability $\pi(a|s)=\frac{1}{2}$. 
	
	The battery state $b^\prime<b$ is accessible from $b$ with a positive probability (w.p.p.) by realizing the action $a=1$ for $(b-b^\prime)$ time slots, and  $b^\prime \geq b$ is accessible from $b$ w.p.p. by realizing the action $a=0$ for $(b^\prime-b)$ slots. As the system reaches the battery state $b^\prime$, regardless of the next actions, the battery state remains unchanged with a positive probability. Therefore, we assume the battery state $b^\prime$ for the remainder of the proof.
	
	Let us define $a_D=\mathds{1}\left(D\in\{\Delta_1^\prime,\Delta_2^\prime,\ldots,\Delta_K^\prime\} \right)$ and $n_D=\max \left\{1, \sum_{k=1}^{K}{\mathds{1}\left(\Delta_k^\prime=D\right)}\right\}$, where $\mathds{1}(\cdot)$ is the indicator function, and $a_D=1$ indicates that the Version AoI at $n_D$ nodes is equal to $D$. Otherwise, when no node has a Version age equal to $D$, $a_D$ and $n_D$ become 0 and 1, respectively. For the parameters $0<\beta<1$, $0<p_t<1$, $0\leq\lambda_k<1$, and $0<q_k<1$, where $k\in\{1,2,\ldots,K\}$, realization of actions $a_{\Delta_{max}}$ for $n_{\Delta_{max}}$ slots, $a_{\Delta_{max}-1}$ for $n_{\Delta_{max}-1}$ slots, ..., $a_{1}$ for $n_{1}$ slots, and $a_{0}$ for $n_{0}$ slots will result the state $s^\prime=\left(b^\prime,\Delta_1^\prime,\Delta_2^\prime,\ldots,\Delta_K^\prime,\Delta_C^\prime\right)$ for the system w.p.p.. Recall that $\Delta_C^\prime=\min\{\Delta_1^\prime,\Delta_2^\prime,\ldots,\Delta_K^\prime\}$.
	This statement is true for $b^\prime>0$. When $b^\prime=0$, realizing the same sequence of actions with $b^\prime=1$ results in the same $s^\prime$, knowing that the last non-zero action $a_D$ reduces $b^\prime$ to $0$, with a positive probability. Therefore, any non-transient state $s^\prime$ is accessible from any other non-transient $s$. In the following, we consider the cases with $p_t=1$ and $\lambda_i=1, i\in\{1,2,\ldots,K\}$, separately.
	
	In case of $p_t=1$, the sequence of actions to reach $s^\prime$ w.p.p. will simply be $a_{\Delta_{max}}, a_{\Delta_{max}-1},$ $\ldots,a_2$, and $a_1$. In this case, the transient set further includes the states in which more than one node has the Version age $1$, more than $3$ nodes have the Version age $2$, ..., and more than $2n-1$ nodes have the Version age $n$, for $n<\Delta_{max}$.
	
	When $\lambda_{i+1}=1, i\in\{1,2,\ldots,K\}$, node $i+1$ is always updated by node $i$ if it has a newer version than node $i+1$. Therefore, state $s=\left(b,\Delta_1^\prime,\Delta_2^\prime,\ldots,\Delta_i^\prime,\Delta_{i+1}^\prime,\ldots,\Delta_K^\prime\right)$ is a transient state, if
	$\Delta_{i+1}^\prime>\Delta_{i}^\prime$ and $\Delta_{i}^\prime \neq \Delta_C$ hold. Otherwise, when $\Delta_{i}^\prime=\Delta_C^\prime=\min\left\{\Delta_1^\prime,\Delta_2^\prime,\ldots,\Delta_K^\prime\right\}$, $s$ is non-transient for all $\Delta_{i+1}^\prime\in\{0,1,\ldots,\Delta_{max}\}$, the reason is that gossiping occurs with a time slot lag. In both cases, the sequence of actions to reach $s^\prime$ w.p.p. is the same as previously mentioned, except when $\Delta_{i}^\prime=\Delta_C^\prime$, where the final realized action updates the node $i$. \hfill $\blacksquare$

	\section{Proof of Theorem 1}
	\label{Appen1}
	
	\noindent The Bellman equation at state $s\!=\!\left(b,\Delta_1,\Delta_2,...,\Delta_K,\Delta_C\right)$ can be simplified as follows:
	
	{\centering \begin{minipage}{\linewidth} {\small
				\begin{align}
					J^\ast+V(s)&=\min_{a\in\left\{0,1\right\}}{\left\{\Delta_{AVG}(s)+\sum_{s^\prime\in S}\ P\left[s^\prime|s,a\right]V(s^\prime)\right\}} \notag \\
					&=\Delta_{AVG}(s)+\min_{a\in\left\{0,1\right\}}{\left\{\sum_{s^\prime\in S} P\left[s^\prime|s,a\right]V(s^\prime)\right\}}.
				\end{align}%
	} \end{minipage} }

	Therefore, the optimal action can be obtained by:
	{\centering \begin{minipage}{\linewidth} {\small
				\begin{align}
					a^\ast(s)\!=\!\argmin_{a\in\left\{0,1\right\}}{\left\{\sum_{s^\prime\in S}\!P\left[s^\prime|s,a\right]V(s^\prime)\right\}}\!=\!
					\begin{cases}
						0, & \Delta V(s)\geq0,\\
						1, & \Delta V(s)<0,\\
					\end{cases}
				\end{align}%
	} \end{minipage} }
	
	\noindent where we have defined $V^0(s)=\sum_{s^\prime\in S}P\left[s^\prime|s,a=0\right]V(s^\prime)$, $V^1(s)\!=\!\sum_{s^\prime\in S}\!P\left[s^\prime|s,a\!=\!1\right]V(s^\prime)$, and $\Delta V(s)\!=\!V^1(s)\!-\!V^0(s)$.
	
	As can be seen, the optimal action $a^\ast(s)$ is related to the sign of $\Delta V(s)$. In what follows, we show that $\Delta V(s)$ and so $a^\ast(s)$ are not a function of $(\Delta_1,\Delta_2,\ldots,\Delta_K)$ values. Using the transition probabilities, we can derive an expression for $\Delta V(s)$ as follows.
	
	\textit{Case 1.} $s$ is a causal state, and the battery is empty, i.e., $b=0$.
	
	{\centering \begin{minipage}{\linewidth} {\small
				\begin{align}
					V^1(s)&=V^0(s)=\sum_{\ \mathbf{g},e,z\ }{V\left(b+e,\Delta_{g_1},...,\Delta_{g_K},\Delta_C+z\right)P_{ze\mathbf{g}r_0}} \notag \\ 
					&+\sum_{\substack{r_i\in\mathcal{R}_1 \\ \mathbf{g},e,z }}{V\!\left(b\!+\!e,\Delta_{g_1},...,\underbrace{\Delta_C\!+\!z}_{node \ i},...,\Delta_{g_K},\Delta_C\!+\!z\right)\!P_{ze\mathbf{g}r_i}},
				\end{align}
	} \end{minipage} }
	
	\noindent where $\mathcal{R}_1=\mathcal{R}\setminus\left\{r_0\right\}$ and $P_{ze\mathbf{g}r_i}=P_zP_eP_\mathbf{g}P_{r_i}$. Therefore, $\Delta V(s)=0$ $\forall s\in S$, and the action $a=0$ is optimal for all $g_k$ and $\left(\Delta_1,\Delta_2,\ldots,\Delta_K\right)$ values.
	
	\textit{Case 2.} $s$ is a causal state, and the battery is not empty, i.e., $b>0$.
	
	{\centering \begin{minipage}{\linewidth} {\small
				\begin{align}
					V^0(s)&=\sum_{\ \mathbf{g},e,z\ }{V\left(b+e,\Delta_{g_1},...,\Delta_{g_K},\Delta_C+z\right)P_{ze\mathbf{g}r_0}}\\ 
					&+\!\sum_{\substack{r_i\in\mathcal{R}_1 \\ \mathbf{g},e,z }}{\!V\!\left(b\!+\!e,\Delta_{g_1},...,\underbrace{\Delta_C\!+\!z}_{node  \ i},...,\Delta_{g_K},\Delta_C\!+\!z\right)\!P_{ze\mathbf{g}r_i}} \notag \\
					V^1(s)&=\sum_{\ \mathbf{g},e,z\ }{V\left(b+e,\Delta_{g_1},...,\Delta_{g_K},\Delta_C+z\right)P_{ze\mathbf{g}r_0}}\\
					&+\!\sum_{\substack{r_i\in\mathcal{R}_1 \\ \mathbf{g},e,z }}{V\!\left(b-1+e,\Delta_{g_1},...,\underbrace{z}_{node \ i},...,\Delta_{g_K},z\right)P_{ze\mathbf{g}r_i}} \notag \\
					\Delta V(s)&=\sum_{\substack{r_i\in\mathcal{R}_1 \\ \mathbf{g},e,z }}\!\Big[V\left(b-1+e,\Delta_{g_1},...,z,...,\Delta_{g_K},z\right)\\
					&-V\left(b+e,\Delta_{g_1},...,\Delta_C+z,...,\Delta_{g_K},\Delta_C+z\right)\Big]P_{ze\mathbf{g}r_i}, \notag
				\end{align}
	} \end{minipage} }
	
	\noindent where $\Delta_{g_k}\!=\!g_k\min{\left\{\Delta_k,\Delta_{k-1}\right\}}\!+\!\left(1\!-\!g_k\right)\Delta_k\!+\!z\geq\Delta_C\!+\!z$. In Lemma 2 we prove that the expression:
	
	{\centering \begin{minipage}{\linewidth} {\small
				\begin{align}
					\label{DV_Expression}
					V&\!\left(b\!-\!1\!+\!e,\Delta_{g_1},...,\Delta_{g_{i-1}},z,\Delta_{g_{i+1}},...,\Delta_{g_K},z\right)\\
					&-V\!\left(b\!+\!e,\Delta_{g_1},...,\Delta_{g_{i-1}},\Delta_C\!+\!z,\Delta_{g_{i+1}},...,\Delta_{g_K},\Delta_C\!+\!z\right) \notag
				\end{align}
	} \end{minipage} }
	
	\noindent is a function of $b$ and $\Delta_C$, but not $\Delta_{g_k}$. Therefore, $\Delta V(s)$ is not a function $(\Delta_1,\Delta_2,\ldots,\Delta_K)$ values, and the proof is complete.
	\hfill $\blacksquare$
	\\
	
	\textbf{Lemma 2.} \textit{In a causal state where $b\geq1$ is the state of the battery and $\delta_k$ is the Version age at the node $k\in\{1,2,\ldots,K\}$, where $\delta_k\geq\delta_C+z_0$ for a fixed number $z_0\in\{0,1,2,\ldots\}$, the following equation is true:}
	
	{\centering \begin{minipage}{\linewidth} {\small
				\begin{align}
					&V\!\left(b\!-\!1,\Delta_1,...,\Delta_K,z_0\right)\!-\!V\!\left(b,\Delta_1^\prime,...,\Delta_K^\prime,\delta_C\!+\!z_0\right) \\
					&=V\!\left(b\!-\!1,\Delta_1^{\prime\prime},...,\Delta_K^{\prime\prime},z_0\right)
					\!-\!V\!\left(b,\delta_C\!+\!z_0,...,\delta_C\!+\!z_0,\delta_C\!+\!z_0\right),  \notag
				\end{align}
	} \end{minipage} }
	
	\noindent where:
	
	{\centering \begin{minipage}{\linewidth} {\small
				\begin{align}
					\begin{matrix}
						\Delta_k\!=\!
						\begin{cases}
							z_0 & k\!\in\!I, \\
							\delta_k & k\!\notin\!I,
						\end{cases}
						&
						\Delta_k^\prime\!=\!
						\begin{cases}
							\delta_C\!+\!z_0 & k\!\in\!I, \\
							\delta_k & k\!\notin\!I,
						\end{cases} \\
						\Delta_k^{\prime\prime}\!=\!
						\begin{cases}
							z_0 & k\!\in\!I, \\
							\delta_C\!+\!z_0 & k\!\notin\!I,
						\end{cases} & I\subseteq\{1,2,\ldots,K\}.
					\end{matrix}
				\end{align}
	} \end{minipage} }
	
	$I$ is an arbitrary subset of nodes. For a particular case in which $I=\{i\}$, it becomes:
	
	{\centering \begin{minipage}{\linewidth} {\small
				\begin{align}
					&V\!\left(b\!-\!1,\delta_1,\ldots,\delta_{i-1},z_0,\delta_{i+1},\ldots,\delta_K,z_0\right)\! \notag \\
					&\quad -\!V\left(b,\delta_1,\ldots,\delta_{i-1},\delta_C\!+\!z_0,\delta_{i+1},\ldots,\delta_K,\delta_C\!+\!z_0\right) \notag \\
					&\!=\!V\!\bigg(\!b\!-\!1,\delta_C\!+\!z_0,\ldots,\delta_C\!+\!z_0,\underbrace{z_0}_{\text{node\ i}},\delta_C\!+\!z_0,\ldots,\delta_C\!+\!z_0,z_0\!\bigg)\! \notag\\
					&\quad -\!V\!\left(b,\delta_C\!+\!z_0,\ldots,\delta_C\!+\!z_0,\delta_C\!+\!z_0\!\right)\!.
				\end{align}
	} \end{minipage} }
	
	In other words, in the left-hand side expression, $\delta_k$ can be replaced by its minimum value, $\delta_C+z_0$.
	
	\textit{Proof:} We use the Value Iteration Algorithm (VIA) to prove the lemma. In this algorithm, the iteration at time $t$ updates the value function as follows:
	
	{\centering \begin{minipage}{\linewidth} {\small
				\begin{align}
					V_{t+1}(s)\!=\!\min_{a\in\left\{0,1\right\}}{\left\{\!\Delta_{AVG}(s)\!+\!\sum_{s^\prime\in S}{P\left[s^\prime|s,a\right]V_t(s^\prime)}\right\}} \ \forall s\in S.
				\end{align}
	} \end{minipage} }
	
	Regardless of the initial value of $V_0(s)$, VIA converges to the value function of the Bellman equation, i.e., $\lim_{t\rightarrow\infty}{V_t(s)}=V(s)\ \forall s\in S$. Therefore, it is sufficient to prove the lemma for $V_t(s)$ for $t\in\{0,1,2,\ldots\}$, i.e.:
	
	{\centering \begin{minipage}{\linewidth} {\small
				\begin{align}
					\label{Lemma1_Eq}
					&V_t\!\left(b\!-\!1,\delta_1,\ldots,\delta_{i-1},z_0,\delta_{i+1},\ldots,\delta_K,z_0\right)\! \notag \\
					&\quad -V_t\left(b,\delta_1,\ldots,\delta_{i-1},\delta_C\!+\!z_0,\delta_{i+1},\ldots,\delta_K,\delta_C\!+\!z_0\right) \notag \\
					&=V_t\left(b\!-\!1,\delta_C\!+\!z_0,\ldots,\delta_C\!+\!z_0,z_0,\delta_C\!+\!z_0,\ldots,\delta_C\!+\!z_0,z_0\right)\! \notag \\
					&\quad -V_t\left(b,\delta_C\!+\!z_0,\ldots,\delta_C\!+\!z_0,\delta_C\!+\!z_0\!\right)\!.
				\end{align}
	} \end{minipage} }
	
	Without loss of generality, we assume that $V_0(s)\!=\!0,\ \forall s\!\in\!S$. Thus, (\ref{Lemma1_Eq}) is true for $t=0$ and for all $I$. Next, assuming it holds till $t>0$, we show it will hold for $t+1$.
	
	\noindent \textit{Case 1.} $b=1$.
	
	{\centering \begin{minipage}{\linewidth} {\small
				\begin{align}
					\vphantom{\bigg\{} &V_{t+1}\left(b\!-\!1,\Delta_1,...,\Delta_K,z_0\right) \\
					&\vphantom{\bigg\{} =\frac{1}{K}\left(\Delta_1\!+\!\Delta_2\!+\!...\!+\!\Delta_K\right) \notag \\
					&+\sum_{\ \mathbf{g},e,z\ }{\!V_t\left(b\!-\!1\!+\!e,\Delta_{g_1},...,\Delta_{g_K},z_0\!+\!z\right)P_{ze\mathbf{g}r_0}} \notag \\
					&\vphantom{\bigg\{}+\sum_{\substack{r_j\in\mathcal{R}_1 \\ \mathbf{g},e,z }}{\!V_t\bigg(b\!-\!1\!+\!e,\Delta_{g_1},...,\underbrace{z_0\!+\!z}_{node\ j},...,\Delta_{g_K},z_0\!+\!z\bigg)P_{ze\mathbf{g}r_j}}, \notag
				\end{align}
				\begin{align}
					\vphantom{\bigg\{} &V_{t+1}\left(b,\Delta_1^\prime,...,\Delta_K^\prime,\delta_C+z_0\right) \\
					&\vphantom{\bigg\{} =\frac{1}{K}\left(\Delta_1^\prime+...+\Delta_K^\prime\right)\! \notag \\
					&\vphantom{\bigg\{} +\sum_{\ \mathbf{g},e,z\ }V_t\big(b\!+\!e,\Delta_{g_1}^\prime,...,\Delta_{g_K}^\prime,\delta_C\!+\!z_0+z\big)\!P_{ze\mathbf{g}r_0}+\min \Bigg\{ \notag \\
					&\vphantom{\bigg\{} \hphantom{=} \sum_{\substack{r_j\in\mathcal{R}_1 \\ \mathbf{g},e,z }}\!V_t\!\bigg(\!b\!+\!e,\Delta_{g_1}^\prime,...,\underbrace{\delta_C\!+\!z_0\!+\!z}_{node\ j},...,\Delta_{g_K}^\prime,\delta_C\!+\!z_0\!+\!z\!\bigg)\!P_{ze\mathbf{g}r_j} , \notag \\
					&\vphantom{\bigg\{} \hphantom{=} \sum_{\substack{r_j\in\mathcal{R}_1 \\ \mathbf{g},e,z }}V_t\bigg(b\!-\!1\!+\!e,\Delta_{g_1}^\prime,...,\underbrace{z}_{node\ j},...,\Delta_{g_K}^\prime,z\bigg)\!P_{ze\mathbf{g}r_j}\Bigg\} \notag \\
					&\vphantom{\bigg\{} =\frac{1}{K}\left(\Delta_1^\prime+...+\Delta_K^\prime\right) \notag \\
					&\vphantom{\bigg\{} +\sum_{\ \mathbf{g},e,z\ }V_t\bigg(b\!+\!e,\Delta_{g_1}^\prime,...,\Delta_{g_K}^\prime,\delta_C\!+\!z_0\!+\!z\bigg)\!P_{ze\mathbf{g}r_0} \notag \\
					&\vphantom{\bigg\{}  +\sum_{\substack{r_j\in\mathcal{R}_1 \\ \mathbf{g},e,z }}V_t\bigg(b\!+\!e,\Delta_{g_1}^\prime,...,\underbrace{\delta_C\!+\!z_0\!+\!z}_{node\ j},...,\Delta_{g_K}^\prime,\delta_C\!+\!z_0\!+\!z\bigg)\!P_{ze\mathbf{g}r_j} \notag \\
					&\vphantom{\bigg\{} +\min\Bigg\{0, \sum_{\substack{r_j\in\mathcal{R}_1 \\ \mathbf{g},e,z }}\bigg[V_t\Big(b\!-\!1\!+\!e,\Delta_{g_1}^\prime,...,\underbrace{z}_{node\ j},...,\Delta_{g_K}^\prime,z\Big) \notag \\
					&\vphantom{\bigg\{} -V_t\Big(b\!+\!e,\Delta_{g_1}^\prime,...,\underbrace{\delta_C\!+\!z_0\!+\!z}_{node\ j},...,\Delta_{g_K}^\prime,\delta_C\!+\!z_0\!+\!z\Big)\bigg]\!P_{ze\mathbf{g}r_j}\Bigg\}, \notag
				\end{align}
	} \end{minipage} }
	
	\noindent where:
	
	{\centering \begin{minipage}{\linewidth} {\small
				\begin{align}
					\Delta_{g_k}\!&=g_k\min{\left\{\Delta_{k-1},\Delta_k\right\}}+\left(1-g_k\right)\Delta_k+z \notag \\
					&=
					\begin{cases}
						z_0+z, & k\in I, \\
						g_kz_0+\left(1-g_k\right)\delta_k+z, & k-1\in I, \\
						g_k\min{\left\{\delta_{k-1},\delta_k\right\}}\!+\!\left(1\!-\!g_k\right)\delta_k\!+\!z, & \text{otherwise}.
					\end{cases}
				\end{align}
				\begin{align}
					\Delta_{g_k}^\prime\!&=g_k\min{\left\{\Delta_{k-1}^\prime\ ,\Delta_k^\prime\right\}}\!+\!\left(1\!-\!g_k\right)\Delta_k^\prime\!+\!z\! \notag \\
					&=\!
					\begin{cases}
						\delta_C+z_0+z, & k\in I, \\
						g_k\left(\delta_C+z_0\right)+\left(1-g_k\right)\delta_k+z, & k\!-\!1\in I, \\
						g_k\min{\left\{\delta_{k-1},\delta_k\right\}}+\left(1\!-\!g_k\right)\delta_k\!+\!z, & \text{otherwise.}
					\end{cases}
				\end{align}
	} \end{minipage} }
	
	We have used $\min{\left\{X,Y\right\}}=X+\min{\left\{0,Y-X\right\}}$. Thus far, we have:
	
	{\centering \begin{minipage}{\linewidth} {\small
				\begin{align}
					\vphantom{\bigg\{}&V_{t+1}\left(b-1,\Delta_1,...,\Delta_K,z_0\right)-V_{t+1}\left(b,\Delta_1^\prime,...,\Delta_K^\prime,\delta_C+z_0\right) \notag \\
					&\vphantom{\bigg\{}=\frac{1}{K}\left[\left(\Delta_1+...+\Delta_K\right)-\left(\Delta_1^\prime+...+\Delta_K^\prime\right)\right]\\
					&\vphantom{\bigg\{}+\sum_{\ \mathbf{g},e,z\ }\big[V_t\left(b-1+e,\Delta_{g_1},...,\Delta_{g_K},z_0+z\right) \notag \\
					&\vphantom{\bigg\{}-V_t\left(b+e,\Delta_{g_1}^\prime,...,\Delta_{g_K}^\prime,\delta_C+z_0+z\right)\big]P_{ze\mathbf{g}r_0} \notag \\
					&\vphantom{\bigg\{}+\sum_{\substack{r_j\in\mathcal{R}_1 \\ \mathbf{g},e,z }}\bigg[V_t\big(b\!-\!1\!+\!e,\Delta_{g_1},...,\underbrace{z_0\!+\!z}_{node\ j},...,\Delta_{g_K},z_0\!+\!z\big) \notag \\
					&\vphantom{\bigg\{}-V_t\big(b\!+\!e,\Delta_{g_1}^\prime,...,\underbrace{\delta_C\!+\!z_0\!+\!z}_{node\ j},...,\Delta_{g_K}^\prime,\delta_C\!+\!z_0\!+\!z\big)\bigg]P_{ze\mathbf{g}r_j} \notag \\
					&\vphantom{\bigg\{}-\min\Bigg\{0, \sum_{\substack{r_j\in\mathcal{R}_1 \\ \mathbf{g},e,z }}\bigg[V_t\Big(b\!-\!1\!+\!e,\Delta_{g_1}^\prime,...,\underbrace{z}_{node\ j},...,\Delta_{g_K}^\prime,z\Big) \notag \\
					&\vphantom{\bigg\{}-V_t\Big(b\!+\!e,\Delta_{g_1}^\prime,...,\underbrace{\delta_C\!+\!z_0\!+\!z}_{node\ j},...,\Delta_{g_K}^\prime,\delta_C\!+\!z_0\!+\!z\Big)\bigg]P_{ze\mathbf{g}r_j}\Bigg\}. \notag
				\end{align}%
	} \end{minipage} }
	
	It is notable that $ \Delta_{g_k}^\prime\!=\!\delta_C\!+\!\Delta_{g_k},\ k\in I^\prime,$ and $ \Delta_{g_k}^\prime\!=\!\Delta_{g_k},\ k\notin I^\prime$, where:
	
	{\centering \begin{minipage}{\linewidth} {\small
				\begin{align}
					I^\prime=\left\{k\middle| k\in I\right\}\cup\left\{k\middle| k-1\in I\ \&\ g_k=1\right\}.
				\end{align}
	} \end{minipage} }
	
	We also have $\Delta_{g_k}^{\prime\prime}=\Delta_{g_k}=z_0\!+\!z,\ k\in I^\prime$, and $\Delta_k^{\prime\prime}\!=\!\delta_C\!+\!z_0\!+\!z,\ k\notin I^\prime$, which is true for:
	
	{\centering \begin{minipage}{\linewidth} {\small
				\begin{align}
					\Delta_{g_k}^{\prime\prime}&=g_k\min{\left\{\Delta_{k-1}^{\prime\prime}\ ,\Delta_k^{\prime\prime}\right\}}\!+\!\left(1\!-\!g_k\right)\Delta_k^{\prime\prime}\!+\!z \notag \\
					&=
					\begin{cases}
						z_0+z, & k\in I, \\
						g_kz_0\!+\!\left(1\!-\!g_k\right)(\delta_C\!+\!z_0)\!+\!z, & k\!-\!1\in I, \\
						\delta_C+z_0+z, & \text{otherwise.}
					\end{cases}
				\end{align}
	} \end{minipage} }
	
	Thus, according to the assumption:
	
	{\centering \begin{minipage}{\linewidth} {\small
				\begin{align}
					&V_t\bigg(b-1,\Delta_{g_1},...,\underbrace{z_0+z}_{node\ j},...,\Delta_{g_K},z_0+z\bigg) \notag \\
					&\quad -V_t\bigg(b,\Delta_{g_1}^\prime,...,\underbrace{\delta_C+z_0+z}_{node\ j},...,\Delta_{g_K}^\prime,\delta_C+z_0+z\bigg) \notag \\
					&=V_t\bigg(\!b\!-\!1,\Delta_{g_1}^{\prime\prime},...,\underbrace{z_0\!+\!z}_{node\ j},...,\Delta_{g_K}^{\prime\prime},z_0\!+\!z\!\bigg)\! \notag \\
					&\quad -V_t\bigg(\!b,\delta_C\!+\!z_0\!+\!z,...,\delta_C\!+\!z_0\!+\!z,\delta_C\!+\!z_0\!+\!z\!\bigg)\!.
				\end{align}
	} \end{minipage} }
	
	Also, we have:
	
	{\centering \begin{minipage}{\linewidth} {\small
				\begin{align}
					&V_t\bigg(b-1,\Delta_{g_1}^\prime,...,\underbrace{z}_{node\ j},...,\Delta_{g_K}^\prime,z\bigg) \notag \\
					&\quad -V_t\bigg(b,\Delta_{g_1}^\prime,...,\underbrace{\delta_C+z_0+z}_{node\ j},...,\Delta_{g_K}^\prime,\delta_C+z_0+z\bigg) \notag \\
					&=V_t\bigg(b-1,\delta_C+z_0+z,...,\underbrace{z}_{node\ j},...,\delta_C+z_0+z,z\bigg) \notag \\
					&\quad -V_t\big(b,\delta_C+z_0+z,...,\delta_C+z_0+z,\delta_C+z_0+z\big).
				\end{align}
	} \end{minipage} }
	
	It can be simply shown that
	$
	\frac{1}{K}\big[\left(\Delta_1+\Delta_2+...+\Delta_K\right)-\left(\Delta_1^\prime+\Delta_2^\prime+...+\Delta_K^\prime\right)\big]=-\frac{n_I}{K}\delta_C$. Thus, we have:
	
	{\centering \begin{minipage}{\linewidth} {\small
				\begin{align}
					&\vphantom{\bigg\{} V_{t+1}\!\left(b\!-\!1,\Delta_1,...,\Delta_K,z_0\right)\!-\!V_{t+1}\left(b,\Delta_1^\prime,...,\Delta_K^\prime,\delta_C\!+\!z_0\right) \notag \\
					&\vphantom{\bigg\{} =-\frac{n_I}{K}\delta_c+\sum_{\ \mathbf{g},e,z\ }\big[V_t\left(b-1+e,\Delta_{g_1}^{\prime\prime},...,\Delta_{g_K}^{\prime\prime},z_0+z\right) \notag \\
					&\vphantom{\bigg\{} -V_t\left(b+e,\delta_C+z_0+z,...,\delta_C+z_0+z\right)\big]P_{ze\mathbf{g}r_0} \notag \\
					&\vphantom{\bigg\{} +\sum_{\substack{r_j\in\mathcal{R}_1 \\ \mathbf{g},e,z }}\!\bigg[V_t\big(b-1+e,\Delta_{g_1}^{\prime\prime},...,z_0+z,...,\Delta_{g_K}^{\prime\prime},z_0+z\big) \notag \\
					&\vphantom{\bigg\{} -V_t\big(b\!+\!e,\delta_C\!+\!z_0\!+\!z,...,\delta_C\!+\!z_0\!+\!z\big)\bigg]P_{ze\mathbf{g}r_j}-\min\Bigg\{0, \notag \\
					&\vphantom{\bigg\{} \hphantom{=} \sum_{\substack{r_j\in\mathcal{R}_1 \\ \mathbf{g},e,z }}\!\bigg[V_t\big(b\!-\!1\!+\!e,\delta_C\!+\!z_0\!+\!z,...,\underbrace{z}_{\text{node\ j}},...,\delta_C\!+\!z_0\!+\!z,z\big) \notag \\
					&\vphantom{\bigg\{} -V_t\big(b+e,\delta_C+z_0+z,...,\delta_C+z_0+z\big)\bigg]P_{ze\mathbf{g}r_j}\Bigg\} \notag \\
					&\vphantom{\bigg\{} =\frac{1}{K}\left[\left(\Delta_1^{\prime\prime}+...+\Delta_K^{\prime\prime}\right)-\left(\delta_C+z_0+...+\delta_C+z_0\right)\right] \notag \\
					&\vphantom{\bigg\{} +\sum_{\ \mathbf{g},e,z\ }{V_t\left(b-1+e,\Delta_{g_1}^{\prime\prime},...,\Delta_{g_K}^{\prime\prime},z_0+z\right)P_{ze\mathbf{g}r_0}} \notag \\
					&\vphantom{\bigg\{} +\sum_{\substack{r_j\in\mathcal{R}_1 \\ \mathbf{g},e,z }}{V_t\left(b-1+e,\Delta_{g_1}^{\prime\prime},...,z_0+z,...,\Delta_{g_K}^{\prime\prime},z_0+z\right)P_{ze\mathbf{g}r_j}} \notag \\
					&\vphantom{\bigg\{} -\sum_{\ \mathbf{g},e,z\ }{V_t\left(b+e,\delta_C+z_0+z,...,\delta_C+z_0+z\right)P_{ze\mathbf{g}r_0}} \notag \\
					&\vphantom{\bigg\{} -\sum_{\substack{r_j\in\mathcal{R}_1 \\ \mathbf{g},e,z }}{V_t\left(b\!+\!e,\delta_C\!+\!z_0\!+\!z,...,\delta_C\!+\!z_0\!+\!z\right)P_{ze\mathbf{g}r_j}}\!-\min\Bigg\{0, \notag \\
					&\vphantom{\bigg\{} \hphantom{=} \sum_{\substack{r_j\in\mathcal{R}_1 \\ \mathbf{g},e,z }}\bigg[V_t\big(b\!-\!1\!+\!e,\delta_C\!+\!z_0\!+\!z,...,\underbrace{z}_{\text{node\ j}},...,\delta_C\!+\!z_0\!+\!z,z\big) \notag \\
					&\vphantom{\bigg\{} -V_t\big(b+e,\delta_C+z_0+z,...,\delta_C+z_0+z\big)\bigg]P_{ze\mathbf{g}r_j}\Bigg\} \notag 
				\end{align}
	} \end{minipage} }
	
	{\centering \begin{minipage}{\linewidth} {\small
				\begin{align}
					&\vphantom{\bigg\{} =\frac{1}{K}\left(\Delta_1^{\prime\prime}+...+\Delta_K^{\prime\prime}\right) \notag \\
					&\vphantom{\bigg\{} +\sum_{\ \mathbf{g},e,z\ }{\!V_t\left(b\!-\!1\!+\!e,\Delta_{g_1}^{\prime\prime},...,\Delta_{g_K}^{\prime\prime},z_0\!+\!z\right)\!P_{ze\mathbf{g}r_0}} \notag \\
					&\vphantom{\bigg\{} +\sum_{\substack{r_j\in\mathcal{R}_1 \\ \mathbf{g},e,z }}{\!V_t\left(b\!-\!1\!+\!e,\Delta_{g_1}^{\prime\prime},...,z_0\!+\!z,...,\Delta_{g_K}^{\prime\prime},z_0\!+\!z\right)\!P_{ze\mathbf{g}r_j}} \notag \\
					&\vphantom{\bigg\{} -\frac{1}{K}\left(\delta_C+z_0+...+\delta_C+z_0\right) \notag \\
					&\vphantom{\bigg\{} -\sum_{\ \mathbf{g},e,z\ }{V_t\left(b+e,\delta_C+z_0+z,...,\delta_C+z_0+z\right)P_{ze\mathbf{g}r_0}} \notag \\
					&\vphantom{\bigg\{} -\min\Bigg\{\sum_{\substack{r_j\in\mathcal{R}_1 \\ \mathbf{g},e,z }}{V_t\left(b+e,\delta_C+z_0+z,...,\delta_C+z_0+z\right)P_{ze\mathbf{g}r_j}}, \notag \\
					&\vphantom{\bigg\{} \hphantom{=} \sum_{\substack{r_j\in\mathcal{R}_1 \\ \mathbf{g},e,z }}\!V_t\big(b\!-\!1\!+\!e,\delta_C\!+\!z_0\!+\!z,...,\underbrace{z}_{\text{node\ j}},...,\delta_C\!+\!z_0\!+\!z,z\big)P_{ze\mathbf{g}r_j}\Bigg\} \notag \\
					&\vphantom{\bigg\{} =V_{t+1}\!\left(b\!-\!1,\Delta_1^{\prime\prime},...,\Delta_K^{\prime\prime},z_0\right)\!-\!V_{t+1}\!\left(b,\delta_C\!+\!z_0,...,\delta_C\!+\!z_0,\delta_C\!+\!z_0\right)
				\end{align}
	} \end{minipage} }
	
	\textit{Case 2.} $b>1$:
	
	{\centering \begin{minipage}{\linewidth} {\small
				\begin{align}
					&\vphantom{\bigg\{} V_{t+1}\left(b-1,\Delta_1,...,\Delta_K,z_0\right)-V_{t+1}\left(b,\Delta_1^\prime,...,\Delta_K^\prime,\delta_C+z_0\right) \notag \\
					&\vphantom{\bigg\{} =\frac{1}{K}\left[(\Delta_1+...+\Delta_K)-(\Delta_1^\prime+...+\Delta_K^\prime)\right] \\
					&\vphantom{\bigg\{} +\sum_{\ \mathbf{g},e,z\ }\big[V_t\left(b-1+e,\Delta_{g_1},...,\Delta_{g_K},z_0+z\right) \notag \\
					&\vphantom{\bigg\{} -V_t\left(b+e,\Delta_{g_1}^\prime,...,\Delta_{g_K}^\prime,\delta_C+z_0+z\right)\big]P_{ze\mathbf{g}r_0} \notag \\
					&\vphantom{\bigg\{} +\sum_{\substack{r_j\in\mathcal{R}_1 \\ \mathbf{g},e,z }}\bigg[V_t\big(b-1+e,\Delta_{g_1},...,\underbrace{z_0+z}_{node\ j},...,\Delta_{g_K},z_0+z\big) \notag \\
					&\vphantom{\bigg\{} -V_t\big(b+e,\Delta_{g_1}^\prime,...,,\underbrace{\delta_C+z_0+z}_{node\ j},...,\Delta_{g_K}^\prime,\delta_C\!+\!z_0\!+\!z\big)\bigg]P_{ze\mathbf{g}r_j} \notag \\
					&\vphantom{\bigg\{} +\min\Bigg\{0, \sum_{\substack{r_j\in\mathcal{R}_1 \\ \mathbf{g},e,z }}\bigg[V_t\big(b-2+e,\Delta_{g_1},...,\underbrace{z}_{node\ j},...,\Delta_{g_K},z\big) \notag \\
					&\vphantom{\bigg\{} -V_t\big(b-1+e,\Delta_{g_1},...,\underbrace{z_0+z}_{node\ j},...,\Delta_{g_K},z_0+z\big)\bigg]P_{ze\mathbf{g}r_j}\Bigg\} \notag \\
					&\vphantom{\bigg\{} -\min\Bigg\{0, \sum_{\substack{r_j\in\mathcal{R}_1 \\ \mathbf{g},e,z }}\bigg[V_t\big(b-1+e,\Delta_{g_1}^\prime,...,\underbrace{z}_{node\ j},...,\Delta_{g_K}^\prime,z\big) \notag \\
					&\vphantom{\bigg\{} -V_t\big(b\!+\!e,\Delta_{g_1}^\prime,...,\underbrace{\delta_C\!+\!z_0\!+\!z}_{node\ j},...,\Delta_{g_K}^\prime,\delta_C\!+\!z_0\!+\!z\big)\bigg]P_{ze\mathbf{g}r_j}\Bigg\}. \notag 
				\end{align}
	} \end{minipage} }
	
	Using the same calculations as case 1, It can be shown that:
	
	{\centering \begin{minipage}{\linewidth} {\small
				\begin{align}
					&V_{t+1}\!\left(b\!-\!1,\Delta_1,...,\Delta_K,z_0\right)\!-\!V_{t+1}\left(b,\Delta_1^\prime,...,\Delta_K^\prime,\delta_C\!+\!z_0\right)\\
					&\!=\!V_{t+1}\!\left(b\!-\!1,\Delta_1^{\prime\prime},...,\Delta_K^{\prime\prime},z_0\right)\!-\!V_{t+1}\left(b,\delta_C\!+\!z_0,...,\delta_C\!+\!z_0\right), \notag
				\end{align}
	} \end{minipage} }
	
	\noindent and the proof is complete.
	\hfill $\blacksquare$

	\section{Proof of Lemma 1}
	\label{pLemma1_Appen}
	
	We use the Value Iteration (VI) algorithm to prove the lemma. The iteration at time $t$ updates the value function as follows:
	
	{\centering \begin{minipage}{\linewidth} {\small
				\begin{align}
					V_{t+1}(s)\!=\!\min_{a\in\left\{0,1\right\}}{\left\{\!\Delta_{AVG}(s)\!+\!\sum_{s^\prime\in S}{P\left[s^\prime|s,a\right]V_t(s^\prime)}\right\}} \ \forall s\in S.
				\end{align}
	} \end{minipage} }
	
	VIA converges to the value function of the Bellman equation regardless of the initial value of $V_0(s)$, i.e., $\lim_{t\rightarrow\infty}{V_t(s)}=V(s)\ \forall s\in S$. Therefore, it is sufficient to prove that $\forall\ \Delta_{C1}\le\Delta_{C2}$ and $\forall t \in \{0,1,2,\cdots\}$,
	
	{\centering \begin{minipage}{\linewidth} {\small
				\begin{align}
					\label{Lemma2_t}
					V_t&\left(b,\Delta_1,...,\Delta_{i-1},\Delta_{C1},\Delta_{i+1},...,\Delta_K,\Delta_{C1}\right) \notag \\
					&\le V_t\left(b,\Delta_1,...,\Delta_{i-1},\Delta_{C2},\Delta_{i+1},...,\Delta_K,\Delta_{C2}\right).
				\end{align}
	} \end{minipage} }
	
	Without loss of generality, we assume that $V_0\left(b,\Delta_1,...,\Delta_{i-1},\Delta_C,\Delta_{i+1},...,\Delta_K,\Delta_C\right)=0,\ \forall\Delta_C$. Thus, (\ref{Lemma2_t}) is true for $t=0$. Next, with the assumption that (\ref{Lemma2_t}) holds till $t>0$, we show it holds for $t+1$. Let us define $x$, $y$, $u$, and $w$ as follows:
	
	{\centering \begin{minipage}{\linewidth} {\small
				\begin{align}
					x=\Delta_{AVG}\left(s_1\right)+\sum_{s^\prime\in S}{P\left[s^\prime|s_1,a=0\right]V_t(s^\prime)}, \\
					y=\Delta_{AVG}\left(s_1\right)+\sum_{s^\prime\in S}{P\left[s^\prime|s_1,a=1\right]V_t(s^\prime)}, \\
					u=\Delta_{AVG}\left(s_2\right)+\sum_{s^\prime\in S}{P\left[s^\prime|s_2,a=0\right]V_t(s^\prime)}, \\
					w=\Delta_{AVG}\left(s_2\right)+\sum_{s^\prime\in S}{P\left[s^\prime|s_2,a=1\right]V_t(s^\prime)}, 
				\end{align}%
	} \end{minipage} }
	
	\noindent where,
	
	{\centering \begin{minipage}{\linewidth} {\small
	\begin{gather}
		s_1=(b,\Delta_1,...,\Delta_{i-1},\Delta_{C1},\Delta_{i+1},...,\Delta_K,\Delta_{C1}), \\
		s_2=(b,\Delta_1,...,\Delta_{i-1},\Delta_{C2},\Delta_{i+1},...,\Delta_K,\Delta_{C2}), \\
		\Delta_{AVG}\!\left(s_1\right)=\!\frac{1}{K}\!\left(\Delta_1\!+\!...\!+\!\Delta_{i-1}\!+\!\Delta_{C1}\!+\!\Delta_{i+1}\!+\!...\!+\!\Delta_K\right), \\
		\Delta_{AVG}\!\left(s_2\right)=\!\frac{1}{K}\!\left(\!\Delta_1\!+\!...\!+\!\Delta_{i-1}\!+\!\Delta_{C2}\!+\!\Delta_{i+1}\!+\!...\!+\!\Delta_K\right).
	\end{gather}
} \end{minipage} }
	
	Accordingly, we have $ V_{t+1}\left(s_1\right)\!=\!\min{\left\{x,y\right\}} $ and $ V_{t+1}\left(s_2\right)\!=\!\min{\left\{u,w\right\}}$. 
	
	Using the transition probabilities, we can show that:
	
	{\centering \begin{minipage}{\linewidth} {\small
				\begin{align}
					x&\!=\!\Delta_{AVG}\left(s_1\right)\!+\!\sum_{\mathbf{g},e,z }{\!V_t\!\left(b\!+\!e,\Delta_{g_1},...,\Delta_{g_K},\Delta_{C1}\!+\!z\right)\!P_{ze\mathbf{g}r_0}} \notag \\
					&\!+\!\sum_{\substack{r_j\in\mathcal{R}_1\\ \mathbf{g},e,z}}{\!V_t\!\left(b\!+\!e,\!\Delta_{g_1},\!...,\underbrace{\!\Delta_{C1}\!+\!z}_{node\ j},\!...,\!\Delta_{g_K},\!\Delta_{C1}\!+\!z\right)\!P_{ze\mathbf{g}r_j}},
				\end{align}
				\begin{align}
					u&\!=\!\Delta_{AVG}\left(s_2\right)\!+\!\sum_{\mathbf{g},e,z }{\!V_t\!\left(b\!+\!e,\Delta_{g_1},...,\Delta_{g_K},\Delta_{C2}+z\right)\!P_{ze\mathbf{g}r_0}} \notag \\
					&+\!\sum_{\substack{r_j\in\mathcal{R}_1 \\ \mathbf{g},e,z }}{\!V_t\!\left(b\!+\!e,\!\Delta_{g_1},\!...,\underbrace{\!\Delta_{C2}\!+\!z}_{node\ j},\!...,\!\Delta_{g_K},\!\Delta_{C2}\!+\!z\right)\!P_{ze\mathbf{g}r_j}}.
				\end{align}
	} \end{minipage} }
	
	Recall that $\Delta_{g_i}=g_i\min{\left\{\Delta_i,\Delta_{i-1}\right\}}+\left(1-g_i\right)\Delta_i+z$ and in the case where $\Delta_i=\Delta_C$ we have $\min{\left\{\Delta_i,\Delta_{i-1}\right\}}=\Delta_i=\Delta_C$; thus, $\Delta_{g_i}$ is simplified to $\Delta_{g_i}=\Delta_C+z$. As can be seen, each term in $x$ is smaller than (or equal to) the corresponding term in $u$. Therefore, $x\le u$. Moreover, if $b=0$, then $y=x$ and $w=y$, otherwise, if $b\neq0$ we have:
	
	{\centering \begin{minipage}{\linewidth} {\small
				\begin{align}
					y&\!=\!\Delta_{AVG}\left(s_1\right)\!+\!\sum_{\mathbf{g},e,z }{\!V_t\!\left(b\!+\!e,\Delta_{g_1},...,\Delta_{g_K},\Delta_{C1}\!+\!z\right)P_{ze\mathbf{g}r_0}} \notag\\
					&+\!\sum_{\substack{r_j\in\mathcal{R}_1 \\ \mathbf{g},e,z }}{\!V_t\!\left(b\!-\!1\!+\!e,\!\Delta_{g_1},\!...,\underbrace{\!z}_{node\ j},\!...,\!\Delta_{g_K},\!z\right)P_{ze\mathbf{g}r_j}},
				\end{align}
				\begin{align}
					w&\!=\!\Delta_{AVG}\left(s_2\right)\!+\!\sum_{\mathbf{g},e,z }{\!V_t\!\left(b\!+\!e,\Delta_{g_1},...,\Delta_{g_K},\Delta_{C2}\!+\!z\right)P_{ze\mathbf{g}r_0}} \notag \\
					&+\!\sum_{\substack{r_j\in\mathcal{R}_1 \\ \mathbf{g},e,z }}{\!V_t\!\left(b\!-\!1\!+\!e,\!\Delta_{g_1},\!...,\underbrace{\!z}_{node\ j},\!...,\!\Delta_{g_K},\!z\right)P_{ze\mathbf{g}r_j}}.
				\end{align}
	} \end{minipage} }

	It also can be seen that $y\le w$. Finally, we know that if $x\le u$ and $y\le w$, then $\min{\left\{x,y\right\}}\le\min{\left\{u,w\right\}}$; therefore, $V_{t+1}\left(s_1\right)\le V_{t+1}\left(s_2\right)$ and the proof is complete.
	\hfill $\blacksquare$

	\section{Proof of Theorem 2}
	\label{Appen2}
	
	The optimal action $a^\ast(s)$ is determined by the sign of $\Delta V(s)$, which is a weighted sum of (\ref{DV_Expression}) with positive weights. Therefore, $\Delta V(s)$ will be positive if (\ref{DV_Expression}) is positive for all the causal states. In Lemma 1, we proved that the second term in (\ref{DV_Expression}) is increasing in $\Delta_C$. Thus, (\ref{DV_Expression}) decreases with $\Delta_C$, and it can become negative for large enough values of $\Delta_C$, resulting in the action $a=1$.
	\hfill $\blacksquare$

	\section{Analysis of Requesting Node's Index in the State Space}
	\label{AppenNewState}
	By including the index of the requesting node in the state vector, we can define a new state vector as $\hat{s}\!=\!\big(b,\Delta_1,\dots,{\Delta}_K,{\Delta}_C,r\big)\in \hat{S}$, where $r\in\mathcal{R}=\left\{r_0,r_1,...,r_K\right\}$. The transition porbabilities $P\left[\hat{s}^\prime\middle|\hat{s},a\right]$ can be calculated:
	
	{\centering \begin{minipage}{\linewidth} {\small
				\begin{align}
					P\left[\hat{s}^\prime|\hat{s},a\right]=\sum_{\ \mathbf{g},e,z\ } P\left[\hat{s}^\prime|\hat{s},a,\mathbf{g},e,z\right]P_zP_eP_\mathbf{g},
				\end{align}
	} \end{minipage} }
	
	\noindent where ${\hat{s}}^\prime=\big(b^\prime,\Delta_1^\prime,\dots,\Delta_K^\prime,\Delta_C^\prime,r^\prime\big)\in \hat{S}$.
	
	By following the same steps as outlined in section \ref{TransProb}, we obtain the transition porbabilities for different cases:
	
	\begin{itemize}
		\itemindent=-6pt
		\item $r=r_0$:
		
		{\centering \begin{minipage}{\linewidth} {\small
					\begin{align}
						P\!\left[\hat{s}^\prime\middle|\hat{s},a,\mathbf{g},e,z\right]
						\!=\!
						\begin{cases}
							P_{r^\prime} & \parbox{3.5cm}{$b^\prime\!=\!b\!+\!e,\ \Delta_k^\prime\!=\!\Delta_{g_k},\\ \Delta_C^\prime=\!\!\Delta_C\!+\!z,$} \\ \\
							0 & \text{otherwise,}
						\end{cases}
					\end{align}
		} \end{minipage} }
		
		\item $r=r_i$, and $a=0$ or $b=0$:
		
		{\centering \begin{minipage}{\linewidth} {\small
					\begin{align}
						P\!\left[\hat{s}^\prime|\hat{s},a,\mathbf{g},e,z\right]
						\!=\!
						\begin{cases}
							P_{r^\prime} & \parbox{3.5cm}{$b^\prime\!=\!b\!+\!e,\ \Delta_k^\prime\!=\!\Delta_{g_k},\\ \Delta_i^\prime\!=\!\Delta_C^\prime=\Delta_C\!+\!z,$} \\ \\
							0 & \text{otherwise,} 
						\end{cases}
					\end{align}
		} \end{minipage} }
		
		\item $r=r_i$, $a=1$, and $b\geq1$:
		
		{\centering \begin{minipage}{\linewidth} {\small
					\begin{align}
						P\!\left[\hat{s}^\prime|\hat{s},a,\mathbf{g},e,z\right]
						\!=\!
						\begin{cases}
							P_{r^\prime} & \parbox{3.5cm}{$b^\prime\!=\!b\!-\!1\!+\!e,\ \Delta_k^\prime\!=\!\Delta_{g_k},\\ \Delta_i^\prime\!=\!\Delta_C^\prime\!=\!z,$} \\ \\
							0 & \text{otherwise.}
						\end{cases} 
					\end{align}
		} \end{minipage} }
		
	\end{itemize}
	
	The Bellman equation (\ref{Bellman_eqn}) can be simplified to:
	
	{\centering \begin{minipage}{\linewidth} {\small
				\begin{align}
					J^\ast\!+\!V(\hat{s})\!=\!\Delta_{AVG}({\hat{s}})\!+\!\min_{a\in\left\{0,1\right\}}{\left\{\sum_{{\hat{s}}^\prime\in {\hat{S}}} \!P\left[{\hat{s}}^\prime|{\hat{s}},a\right]V({\hat{s}}^\prime)\right\}}.
				\end{align}%
				\begin{align}
					a^\ast({\hat{s}})\!=\!\argmin_{a\in\left\{0,1\right\}}{\left\{\!\sum_{{\hat{s}}^\prime\in {\hat{S}}}\ \!P\left[{\hat{s}}^\prime|{\hat{s}},a\right]V({\hat{s}}^\prime)\!\right\}}\!=\!
					\begin{cases}
						0, & \Delta V({\hat{s}})\geq0,\\
						1, & \Delta V({\hat{s}})<0.\\
					\end{cases}
				\end{align}%
	} \end{minipage} }
	
	As can be seen, the optimal action $a^\ast(\hat{s})$ is related to the sign of $\Delta V(\hat{s})$, where $\Delta V(\hat{s})=V^1(\hat{s})-V^0(\hat{s})$, $V^0(\hat{s})=\sum_{\hat{s}^\prime\in \hat{S}}P\left[\hat{s}^\prime|\hat{s},a=0\right]V(\hat{s}^\prime)$, and $V^1(\hat{s})=\sum_{\hat{s}^\prime\in \hat{S}}P\left[\hat{s}^\prime|\hat{s},a=1\right]V(\hat{s}^\prime)$. We show that $V^0(\hat{s})$, $V^1(\hat{s})$, and $\Delta V(\hat{s})$ have similar structure as $V^(s)$, $V^1(s)$, and $\Delta V(s)$, respectively. Therefore, Theorems 1 and 2 can be proven using the identical approach for the new state space.
	
	\noindent \textbf{\textit{Case 1.}} $b=0$.
	
	\begin{itemize}
		
		\item $r=r_0$:
		
		{\centering \begin{minipage}{\linewidth} {\small
					\begin{align}
						\vphantom{\bigg\{} V^1(\hat{s})&=V^0(\hat{s})\\
						& \vphantom{\bigg\{}=\sum_{\substack{r^\prime\in\mathcal{R} \\ \mathbf{g},e,z }}{V\left(b+e,\Delta_{g_1},...,\Delta_{g_K},\Delta_C+z,r^\prime\right)P_{ze\mathbf{g}r^\prime}}, \notag
					\end{align}%
		} \end{minipage} }
		
		\item $r=r_i$:
		
		{\centering \begin{minipage}{\linewidth} {\small
					\begin{align}
						&\vphantom{\bigg\{} V^1(\hat{s})= V^0(\hat{s})\\ \notag
						&\vphantom{\bigg\{} =\sum_{\substack{r^\prime\in\mathcal{R} \\ \mathbf{g},e,z }}{\!V\!\left(b\!+\!e,\Delta_{g_1},...,\Delta_C\!+\!z,...,\Delta_{g_K},\Delta_C\!+\!z,r^\prime\right)P_{ze\mathbf{g}r^\prime}},
					\end{align}%
		} \end{minipage} }
	\end{itemize}
	
	\noindent so $\Delta V(\hat{s})=\Delta V(s)=0$, and the action $a=0$ is optimal for all $\left(\Delta_1,\Delta_2,\ldots,\Delta_K\right)$ values.
	
	\noindent \textbf{\textit{Case 2.}} $b>0$.
	
	\begin{itemize}
		\item $r=r_0$: same as case $1$, we have $\Delta V(\hat{s})=0$. 
		\item $r=r_i$:
		
		{\centering \begin{minipage}{\linewidth} {\small
					\begin{align}
						&\vphantom{\bigg\{} \Delta V(\hat{s})\\
						&\vphantom{\bigg\{} =\sum_{\substack{r^\prime\in\mathcal{R}_1 \\ \mathbf{g},e,z }}\Big[V\left(b-1+e,\Delta_{g_1},...,z,...,\Delta_{g_K},z,r^\prime\right) \notag\\
						&\vphantom{\bigg\{} -V\left(b+e,\Delta_{g_1},...,\Delta_C+z,...,\Delta_{g_K},\Delta_C+z,r^\prime\right)\Big]P_{ze\mathbf{g}r^\prime}. \notag
					\end{align}
		} \end{minipage} }
	\end{itemize}
	
	We see that $\Delta V(\hat{s})$ has a similar structure to $\Delta V(s)$. Therefore, similar to the proof of Lemma 2, we can establish that the expression: 
	
	{\centering \begin{minipage}{\linewidth} {\small
				\begin{align}
					&V\!\left(b\!-\!1\!+\!e,\Delta_{g_1},...,\Delta_{g_{i-1}},z,\Delta_{g_{i+1}},...,\Delta_{g_K},z,r^\prime\right)\\
					&\!-\!V\!\left(b\!+\!e,\Delta_{g_1},...,\Delta_{g_{i-1}},\!\Delta_C\!+\!z,\Delta_{g_{i\!+\!1}},...,\Delta_{g_K},\Delta_C\!+\!z,r^\prime\right), \notag
				\end{align}
	} \end{minipage} }
	
	\noindent relies on the variables $b$ and $\Delta_C$, while being independent of $\Delta_{g_k}$. Consequently, $\Delta V(\hat{s})$ does not depend on $(\Delta_1,\Delta_2,\ldots,\Delta_K)$, indicating that Theorem 1 holds true within this revised state space. A similar methodology can be explored to establish the validity of Theorem 2 for the new state space.
	
\end{document}